\documentclass[prd,aps,showpacs,epsf,floats,onecolumn]{revtex4}%
\usepackage{amssymb}
\usepackage{amsfonts}
\usepackage{amsmath}
\usepackage{graphicx}%
\providecommand{\U}[1]{\protect\rule{.1in}{.1in}}
\begin{document}
\title{\textbf{From Uncertainty Relations to Quantum Acceleration Limits}}
\author{\textbf{Carlo Cafaro}$^{1,2}$, \textbf{Christian Corda}$^{2,3}$\textbf{,
Newshaw Bahreyni}$^{4}$, \textbf{Abeer Alanazi}$^{1}$}
\affiliation{$^{1}$University at Albany-SUNY, Albany, NY 12222, USA}
\affiliation{$^{2}$SUNY Polytechnic Institute, Utica, NY 13502, USA}
\affiliation{$^{3}$International Institute for Applicable Mathematics and Information
Sciences, B. M. Birla Science Centre, Adarshnagar, Hyderabad 500063, India}
\affiliation{$^{4}$Pomona College, Claremont, CA 91711, USA}

\begin{abstract}
The concept of quantum acceleration limit has been recently introduced for any
unitary time evolution of quantum systems under arbitrary nonstationary
Hamiltonians. While Alsing and Cafaro [Int. J. Geom. Methods Mod. Phys. 21,
2440009 (2024)] used the Robertson uncertainty relation in their derivation,
Pati [arXiv:quant-ph/2312.00864 (2023)] employed the Robertson-Schr\"{o}dinger
uncertainty relation to find the upper bound on the temporal rate of change of
the speed of quantum evolutions. In this paper, we provide a comparative
analysis of these two alternative derivations for quantum systems specified by
an arbitrary finite-dimensional projective Hilbert space. Furthermore,
focusing on a geometric description of the quantum evolution of two-level
quantum systems on a Bloch sphere under general time-dependent Hamiltonians,
we find the most general conditions needed to attain the maximal upper bounds
on the acceleration of the quantum evolution. In particular, these conditions
are expressed explicitly in terms of two three-dimensional real vectors, the
Bloch vector that corresponds to the evolving quantum state and the magnetic
field vector that specifies the Hermitian Hamiltonian of the system. For
pedagogical reasons, we illustrate our general findings for two-level quantum
systems in explicit physical examples characterized by specific time-varying
magnetic field configurations. Finally, we briefly comment on the extension of
our considerations to higher-dimensional physical systems in both pure and
mixed quantum states.

\end{abstract}

\maketitle

\section{Introduction}

Discovering the maximal speed of evolution of a quantum system is an essential
task in quantum information science. The maximal speed is generally achieved
by reducing the time or speeding up the quantum-mechanical process being
considered. We suggest Refs. \cite{frey16,deffner17} for a review on this
topic. The so-called quantum speed limits (QSLs) were historically proposed
for closed systems that evolve in a unitary fashion between orthogonal states
by Mandelstam-Tamm (MT) in Ref. \cite{mandelstam45} and by Margolus-Levitin
(ML) in Ref. \cite{margo98}. The MT bound $\tau_{\mathrm{MT}}\overset
{\text{def}}{=}\pi\hslash/(2\Delta E)$ is a consequence of the Heisenberg
uncertainty relation and is characterized by the variance $\Delta E$ of the
energy of the initial quantum state. The ML bound $\tau_{\mathrm{QSL}}%
\overset{\text{def}}{=}$ $\pi\hslash/\left[  2(E-E_{0})\right]  $, instead, is
based upon the notion of the transition probability amplitude between two
quantum states in Hilbert space and can be described by means of the mean
energy $E$ of the initial state with respect to the ground state energy
$E_{0}$. Clearly, $\hslash$ denotes the reduced Planck constant. For a
presentation of QSL bounds for nonstationary Hamiltonian evolutions between
two arbitrary orthogonal quantum states and to nonunitary evolutions of open
quantum systems, we suggest Refs. \cite{frey16,deffner17}. Finally, for an
investigation on generalizations of the MT quantum speed limit to systems in
mixed quantum states, we suggest \cite{ole22}.

In addition to speed limits, higher-order rates of changes can be equally
relevant in quantum physics
\cite{cai81,cai82,cai84,pati92,pati92b,schot78,eager16}. For instance, the
concept of acceleration in its different forms plays an important role in
adiabatic quantum dynamics \cite{masuda11,masuda22}, in quantum tunneling
dynamics \cite{nakamura16}, and in quantum optimal control employed for fast
cruising throughout large Hilbert spaces \cite{larrouy20}. Beyond simply
considering acceleration, one could also take into account jerk, which is the
time-derivative of acceleration. Jerk plays a crucial role in engineering
applications, particularly in robotics and automation \cite{saridis88}, where
minimizing jerk is a common objective. This is due to the negative impact of
the third derivative of position on control algorithm efficiency
\cite{saridis88}. Furthermore, minimum jerk trajectory techniques
\cite{wise05} have been utilized in quantum settings to optimize paths for
transporting atoms in optical lattices \cite{liu19,matt23}. This is a
significant step in the coherent formation of single ultracold molecules,
which is essential for advancements in quantum information processing and
quantum engineering.

Recently, Pati introduced in Ref. \cite{pati23} the concept of quantum
acceleration limit for unitary time evolution of quantum systems under
nonstationary Hamiltonians. In particular, he proved that the quantum
acceleration is upper bounded by the fluctuation in the derivative of the
Hamiltonian. Inspired by their geometric investigations on the notions of
curvature and torsion of quantum evolutions \cite{paul24A,paul24B}, Alsing and
Cafaro showed in Ref. \cite{alsing24} that the acceleration squared of a
quantum evolution in projective space (for any finite-dimensional quantum
system) is upper bounded by the variance of the temporal rate of change of the
Hamiltonian operator.

In this paper, we have three main novel goals. First, motivated by the
theoretical relevance of these recent findings, we aim to provide a
comparative analysis between these two approaches when applied to any
finite-dimensional quantum system in a pure state. In particular, this
analysis is performed by comparing the conceptual arguments, including the
type of quantum uncertainty relations, used in the derivations of these upper
bounds on the acceleration for quantum evolutions in projective Hilbert space.
Second, to grasp physical insights in simple scenarios, we aim to reformulate
these proofs for the case of single-qubit systems by means of simple vector
algebra techniques equipped with a neat geometric interpretation. Finally, we
aim to present our overarching results for two-level quantum systems through
explicit physical examples defined by particular configurations of
time-varying magnetic fields.

The rest of the paper is organized as follows. In\ Section II, we provide a
comparative analysis between the derivations of the quantum acceleration limit
for arbitrary finite-dimensional quantum systems in a pure state as originally
proposed by Pati \cite{pati23} and Alsing-Cafaro \cite{alsing24}. In Section
III, we discuss the quantum acceleration limit for two-level quantum systems.
In particular, after recasting the acceleration limit in terms of the Bloch
vector $\mathbf{a}$ of the system and of the magnetic field vector
$\mathbf{h}$ that specifies the Hamiltonian of the system, we identify
suitable geometric conditions in terms of $\mathbf{a}$, $\mathbf{h}$, and
their temporal derivatives for which the inequality is saturated and the
quantum acceleration is maximum. In Section IV, we present three illustrative
examples. In the first scenario, the quantum acceleration does not reach the
value of its upper bound (i.e., no saturation). In the second scenario, the
quantum acceleration reaches the value of its upper bound. However, the upper
bound does not assumes its maximum value (i.e., saturation without maximum).
In the third scenario, the quantum acceleration reaches the value of its upper
bound. Moreover, the upper bound achieves its maximum value (i.e., saturation
with maximum). Our concluding remarks appear in Section V. Finally, technical
details are placed in Appendixes A and B.

\section{Quantum acceleration limit}

The uncertainty principle stands as a fundamental and significant outcome of
quantum mechanics. Formulated by Heisenberg \cite{heisenberg} and derived by
Kennard \cite{kennard}, this principle pertains to two conjugate
quantum-mechanical variables and asserts that the precision with which these
variables can be simultaneously measured is constrained by the condition that
the product of the uncertainties in both measurements must be no less than a
value proportional to Planck's constant $\hslash$. The extension of the
uncertainty principle to two arbitrary variables that are not conjugate (i.e.,
arbitrary observables represented by self adjoint operators) was proposed by
Robertson in Ref. \cite{robertson}. Following Robertson's analysis,
Schr\"{o}dinger then provided a generalized version of Robertson's inequality
specified by a tighter constraint relation \cite{schrodinger}. For a unifying
approach to the derivation of uncertainty relations, we suggest Ref.
\cite{englert}.

The uncertainty principle, in one form or another, is at the root of the
derivations of the quantum acceleration limits provided by Pati in Ref.
\cite{pati23} and Alsing-Cafaro in Ref. \cite{alsing24}. In what follows, a
comparative analysis of these two distinct derivations is presented.

Before starting with Pati's derivation, however, we present some basic
ingredients on the geometry of quantum evolution.\textbf{ }Let $\mathcal{H}%
\backslash\left\{  0\right\}  $ be the Hilbert space specified by an $\left(
N+1\right)  $-dimensional complex vector space of normalized state vectors
$\left\{  \left\vert \psi\left(  t\right)  \right\rangle \right\}  $. In
quantum physics, a physical state is not described by a normalized state
vector $\left\vert \psi\left(  t\right)  \right\rangle \in\mathcal{H}%
\backslash\left\{  0\right\}  $. Instead, physical states are specified by a
ray.\ A ray represents the one-dimensional subspace of $\mathcal{H}%
\backslash\left\{  0\right\}  $ given by $\left\{  e^{i\phi\left(  t\right)
}\left\vert \psi\left(  t\right)  \right\rangle :e^{i\phi\left(  t\right)
}\in U(1)\right\}  $, with $\left\vert \psi\left(  t\right)  \right\rangle $
being an element of this subspace. Two quantum state vectors\textbf{
}$\left\vert \psi_{1}\left(  t\right)  \right\rangle $ and $\left\vert
\psi_{2}\left(  t\right)  \right\rangle $ that are elements of the same ray
are equivalent. In\textbf{ }other words, $\left\vert \psi_{A}\left(  t\right)
\right\rangle \simeq\left\vert \psi_{B}\left(  t\right)  \right\rangle $ if
$\left\vert \psi_{A}\left(  t\right)  \right\rangle =e^{i\phi_{AB}\left(
t\right)  }\left\vert \psi_{B}\left(  t\right)  \right\rangle $ for some
$\phi_{AB}\left(  t\right)  \in%
\mathbb{R}
$. The equivalence relation \textquotedblleft$\simeq$\textquotedblright%
\ generates equivalence classes on the $\left(  2N+1\right)  $-dimensional
sphere $S^{2N+1}$. The set of all equivalence classes $S^{2N+1}/U\left(
1\right)  $ determines the space of rays or, equivalently, the space of
physical quantum states. Generally speaking, $S^{2N+1}/U\left(  1\right)  $ is
known as the projective Hilbert space $\mathcal{P}\left(  \mathcal{H}\right)
$. For example, when focusing on the physics of two-level quantum systems, the
two-dimensional Hilbert space $\mathcal{H}_{2}^{1}$ of single-qubit quantum
states (with $N=1$) is viewed in terms of points on the complex projective
Hilbert space $%
\mathbb{C}
P^{1}$ (or, equivalently, the Bloch sphere $S^{3}/U\left(  1\right)
=S^{2}\cong$ $%
\mathbb{C}
P^{1}$ with \textquotedblleft$\cong$\textquotedblright\textbf{\ }denoting
isomorphic spaces) equipped with the Fubini-Study metric (or, equivalently,
the round metric on the sphere). This connection between state vectors in
Hilbert space and rays in projective Hilbert space intermediated by phase
factors can be suitably specified by means of the fiber bundle language
\cite{nakahara}. Loosely speaking, the main elements of a fiber bundle are a
total space $E$, a base space $M$, a fiber space $\mathrm{F}$, a group $G$
that acts on the fibers, and \ a\textbf{ }projection map $\pi$ that projects
the fibers above to points in\textbf{ }$M$. In quantum physics, $\mathcal{H}%
\backslash\left\{  0\right\}  $ acts like $E$, $\mathcal{P}\left(
\mathcal{H}\right)  $ plays the part of $M$, $U\left(  1\right)  $ replicates
$G$, fibers in\textbf{ }$F$ are specified by all unit vectors from the same
ray and, finally, the projection map $\pi$ given by%
\begin{equation}
\pi:\mathcal{H}\backslash\left\{  0\right\}  \ni\left\vert \psi\left(
t\right)  \right\rangle \mapsto\pi\left\vert \psi\left(  t\right)
\right\rangle \overset{\text{def}}{=}\left\vert \psi\left(  t\right)
\right\rangle \left\langle \psi\left(  t\right)  \right\vert \in
\mathcal{P}\left(  \mathcal{H}\right)  \text{,}%
\end{equation}
acts like the projection in the fiber bundle construction. For technical
details on the fiber bundle formalism, we suggest Refs.
\cite{nakahara,eguchi80,bohm91}. For a more in depth discussion on the
geometry of quantum evolutions, we refer to Ref. \cite{crell09}. Finally, for
technical details on the Riemannian structure on manifolds of quantum states
and on the general formalism of (projective) Hilbert spaces, we refer to Refs.
\cite{provost80,mukunda93}.

\subsection{Pati's derivation}

We can now turn to Pati's derivation. Let us recall that the Fubini-Sudy
infinitesimal line element $ds^{2}$ is equal to \cite{anandan90}
\begin{equation}
ds^{2}\overset{\text{def}}{=}4\left[  1-\left\vert \left\langle \psi\left(
t\right)  \left\vert \psi\left(  t+dt\right)  \right.  \right\rangle
\right\vert ^{2}\right]  =\frac{4}{\hslash^{2}}\Delta\mathrm{H}\left(
t\right)  ^{2}dt^{2}\text{,}%
\end{equation}
with $\Delta\mathrm{H}\left(  t\right)  ^{2}\overset{\text{def}}%
{=}\left\langle \psi\left(  t\right)  \left\vert \mathrm{H}\left(  t\right)
^{2}\right\vert \psi\left(  t\right)  \right\rangle -\left\langle \psi\left(
t\right)  \left\vert \mathrm{H}\left(  t\right)  \right\vert \psi\left(
t\right)  \right\rangle ^{2}$ denoting the energy uncertainty of the system,
$i\hslash\partial_{t}\left\vert \psi\left(  t\right)  \right\rangle
=\mathrm{H}\left(  t\right)  \left\vert \psi\left(  t\right)  \right\rangle $
being the time-dependent Schr\"{o}dinger equation, and $\hslash$ representing
the reduced Planck constant. The total distance $s=s(t)$ the system travels on
the projective Hilbert space\textbf{ }is given by%
\begin{equation}
s\left(  t\right)  \overset{\text{def}}{=}\frac{2}{\hslash}\int^{t}%
\Delta\mathrm{H}\left(  t^{\prime}\right)  dt^{\prime}\text{.}%
\end{equation}
Therefore, the speed of transportation $v_{\mathrm{H}}\left(  t\right)  $ of
the quantum system on the projective Hilbert space is equal to%
\begin{equation}
v_{\mathrm{H}}\left(  t\right)  \overset{\text{def}}{=}\frac{ds\left(
t\right)  }{dt}=\frac{2}{\hslash}\Delta\mathrm{H}\left(  t\right)  \text{.}
\label{dust}%
\end{equation}
For clarity, we stress that if one defines the Fubini-Sudy infinitesimal line
element as $ds^{2}\overset{\text{def}}{=}\left[  1-\left\vert \left\langle
\psi\left(  t\right)  \left\vert \psi\left(  t+dt\right)  \right.
\right\rangle \right\vert ^{2}\right]  =\left[  \Delta\mathrm{H}\left(
t\right)  ^{2}/\hslash^{2}\right]  dt^{2}$ \cite{sam94,uzdin12}, the speed of
quantum evolution becomes $v_{\mathrm{H}}\left(  t\right)  \overset
{\text{def}}{=}\Delta\mathrm{H}\left(  t\right)  /\hslash$. Employing
$v_{\mathrm{H}}\left(  t\right)  $ as in Eq. (\ref{dust}), the quantum
acceleration $a_{\mathrm{H}}(t)$ is given by%
\begin{equation}
a_{\mathrm{H}}(t)\overset{\text{def}}{=}\frac{dv_{\mathrm{H}}\left(  t\right)
}{dt}=\frac{2}{\hslash}\frac{d\left[  \Delta\mathrm{H}\left(  t\right)
\right]  }{dt}\text{.} \label{explain}%
\end{equation}
We emphasize that the acceleration $a_{\mathrm{H}}\left(  t\right)  $ is
different from the acceleration $a\left(  t\right)  $ of a quantum particle
\cite{pati92,pati92b}\textbf{. }Indeed, while $a_{\mathrm{H}}\left(  t\right)
$ is the temporal rate of change of the speed of quantum evolution in
projective Hilbert space, $a\left(  t\right)  \overset{\text{def}}%
{=}\left\vert \left\langle d^{2}\hat{x}/dt^{2}\right\rangle \right\vert
=\left\vert dv(t)/dt\right\vert $ where $v(t)\overset{\text{def}}%
{=}\left\langle \hat{v}\right\rangle =\left\langle \psi\left(  0\right)
\left\vert \hat{v}\left(  t\right)  \right\vert \psi\left(  0\right)
\right\rangle =\left\langle \psi\left(  0\right)  \left\vert U^{\dagger
}(t)\hat{v}\left(  0\right)  U(t)\right\vert \psi\left(  0\right)
\right\rangle =\left\langle \psi\left(  t\right)  \left\vert \hat{v}\left(
0\right)  \right\vert \psi\left(  t\right)  \right\rangle $ with
$i\hbar\partial_{t}\left\vert \psi\left(  t\right)  \right\rangle
=$\textrm{\^{H}}$\left(  t\right)  \left\vert \psi\left(  t\right)
\right\rangle $, and $\hat{v}\left(  t\right)  \overset{\text{def}}{=}d\hat
{x}\left(  t\right)  /dt=(i\hbar)^{-1}\left[  \hat{x}\left(  t\right)  \text{,
\textrm{\^{H}}}\left(  t\right)  \right]  $. Clearly $\hat{v}$, $\hat{x}$, and
\textrm{\^{H} }denote the velocity, the position, and the Hamiltonian
operators, respectively. The position operator\textbf{ }$\hat{x}$ is assumed
to be only implicitly time-dependent (i.e\textbf{., }$\partial_{t}$\textbf{
}$\hat{x}=0$\textbf{)} and is understood to be in the Heisenberg picture where
$\hat{x}\left(  t\right)  =U^{\dagger}\left(  t\right)  \hat{x}\left(
0\right)  U(t)$, with $U(t)$ being the unitary evolution operator. As a side
remark, note that $a\left(  t\right)  $ is always positive by definition.
Instead, the sign of $a_{\mathrm{H}}\left(  t\right)  $ can be negative. More
importantly, note that $a_{\mathrm{H}}\left(  t\right)  $ vanishes for
stationary quantum Hamiltonian systems since, in this scenario, the energy
uncertainty of the system does not change in time. Therefore, $a_{\mathrm{H}%
}\left(  t\right)  $ plays a role only for time-varying quantum Hamiltonian
evolutions. Unlike $a_{\mathrm{H}}\left(  t\right)  $, the quantum particle
acceleration $a\left(  t\right)  $ plays a role in both stationary and
non-stationary quantum Hamiltonian evolutions. This is due to the fact that
unlike $\left\langle \mathrm{\hat{H}}\right\rangle $, the\textbf{ }expectation
value $\left\langle \hat{v}\right\rangle $ can depend on time even if the
Hamiltonian operator is constant in time. This, in turn, is a consequence of
the fact that the quantum commutator between $\hat{v}$ and $\mathrm{\hat{H}}$
is generally nonzero. For completeness, we stress that the expectation values
$\left\langle \mathrm{\hat{H}}\right\rangle $ and $\left\langle \hat
{v}\right\rangle $ are taken with respect to a Heisenberg state ket that does
not move with time. To have an intuitive viewpoint on the presence of a
quantum acceleration in the absence of a time-dependent Hamiltonian, we
suggest the following classical scenario. In terms of classical Hamiltonian
mechanics, imagine to have a system with a constant Hamiltonian \textrm{H}%
$_{\mathrm{classical}}$ (i.e., $d$\textrm{H}$_{\mathrm{classical}}/dt=0$)
given by \textrm{H}$_{\mathrm{classical}}\left(  x\text{, }\dot{x}\right)
\overset{\text{def}}{=}(1/2)m\dot{x}^{2}+mgx$. In this classical
case\textbf{,} $a\overset{\text{def}}{=}\ddot{x}=-g\neq0$. Clearly, $m$ and
$g$ denote here the mass of the particle and the acceleration of gravity, respectively.

To prove that the Robertson-Schr\"{o}dinger uncertainty relation provides an
upper bound on the quantum acceleration $a_{\mathrm{H}}(t)$ in\ Eq.
(\ref{explain}), Pati's derivation proceeds as follows. First, we note that
\begin{align}
v_{\mathrm{H}}\left(  t\right)  \frac{dv_{\mathrm{H}}\left(  t\right)  }{dt}
&  =\frac{2}{\hslash}\Delta\mathrm{H}\left(  t\right)  \frac{d}{dt}\left[
\frac{2}{\hslash}\Delta\mathrm{H}\left(  t\right)  \right] \nonumber\\
&  =\frac{2}{\hslash}\sqrt{\left\langle \mathrm{H}^{2}\right\rangle
-\left\langle \mathrm{H}\right\rangle ^{2}}\frac{2}{\hslash}\frac{d}%
{dt}\left[  \sqrt{\left\langle \mathrm{H}^{2}\right\rangle -\left\langle
\mathrm{H}\right\rangle ^{2}}\right] \nonumber\\
&  =\frac{4}{\hslash^{2}}\sqrt{\left\langle \mathrm{H}^{2}\right\rangle
-\left\langle \mathrm{H}\right\rangle ^{2}}\frac{1}{2\sqrt{\left\langle
\mathrm{H}^{2}\right\rangle -\left\langle \mathrm{H}\right\rangle ^{2}}}%
\frac{d}{dt}\left(  \left\langle \mathrm{H}^{2}\right\rangle -\left\langle
\mathrm{H}\right\rangle ^{2}\right) \nonumber\\
&  =\frac{4}{\hslash^{2}}\left(  \frac{\left\langle \mathrm{H\dot{H}%
}+\mathrm{\dot{H}H}\right\rangle }{2}-\left\langle \mathrm{H}\right\rangle
\left\langle \mathrm{\dot{H}}\right\rangle \right) \nonumber\\
&  =\frac{4}{\hslash^{2}}\mathrm{cov}\left(  \mathrm{H}\text{, }%
\mathrm{\dot{H}}\right)  \text{,}%
\end{align}
that is,%
\begin{equation}
v_{\mathrm{H}}\left(  t\right)  a_{\mathrm{H}}(t)=\frac{4}{\hslash^{2}%
}\mathrm{cov}\left(  \mathrm{H}\text{, }\mathrm{\dot{H}}\right)  \text{.}
\label{luo1}%
\end{equation}
The covariance term that appears in Eq. (\ref{luo1}) is formally defined for
two observables $A$ and $B$ as%
\begin{equation}
\mathrm{cov}\left(  A\text{, }B\right)  \overset{\text{def}}{=}\left\langle
\frac{\left(  A-\left\langle A\right\rangle \right)  \left(  B-\left\langle
B\right\rangle \right)  +\left(  B-\left\langle B\right\rangle \right)
\left(  A-\left\langle A\right\rangle \right)  }{2}\right\rangle =\left\langle
\frac{AB+BA}{2}\right\rangle -\left\langle A\right\rangle \left\langle
B\right\rangle \text{.} \label{covariance}%
\end{equation}
The covariance term $\mathrm{cov}\left(  A\text{, }B\right)  $ in Eq.
(\ref{covariance}) plays a key role in the Robertson-Schr\"{o}dinger
uncertainty relation which, for two Hermitian observables $A$ and $B$, is
given by%
\begin{equation}
\Delta A^{2}\Delta B^{2}\geq\mathrm{cov}\left(  A\text{, }B\right)  ^{2}%
+\frac{1}{4}\left\vert \left\langle \left[  A\text{, }B\right]  \right\rangle
\right\vert ^{2}\text{.} \label{SR}%
\end{equation}
A derivation of the inequality in Eq. (\ref{SR}) appears in Appendix A. Then,
taking $A=\mathrm{H}$ and $B=\mathrm{\dot{H}}$, the inequality in Eq.
(\ref{SR}) becomes%
\begin{equation}
\Delta\mathrm{H}^{2}\Delta\mathrm{\dot{H}}^{2}\geq\mathrm{cov}\left(
\mathrm{H}\text{, }\mathrm{\dot{H}}\right)  ^{2}+\frac{1}{4}\left\vert
\left\langle \left[  \mathrm{H}\text{, }\mathrm{\dot{H}}\right]  \right\rangle
\right\vert ^{2}\text{,}%
\end{equation}
that is,%
\begin{equation}
\mathrm{cov}\left(  \mathrm{H}\text{, }\mathrm{\dot{H}}\right)  ^{2}\leq
\Delta\mathrm{H}^{2}\Delta\mathrm{\dot{H}}^{2}-\frac{1}{4}\left\vert
\left\langle \left[  \mathrm{H}\text{, }\mathrm{\dot{H}}\right]  \right\rangle
\right\vert ^{2}\text{.} \label{luo2}%
\end{equation}
Using Eqs. (\ref{dust}), (\ref{luo1}), and (\ref{luo2}), we obtain%
\begin{equation}
\mathrm{cov}\left(  \mathrm{H}\text{, }\mathrm{\dot{H}}\right)  ^{2}%
=\frac{\hslash^{4}}{16}v_{\mathrm{H}}\left(  t\right)  ^{2}a_{\mathrm{H}%
}(t)^{2}=\frac{\hslash^{4}}{16}\frac{4}{\hslash^{2}}\Delta\mathrm{H}%
^{2}a_{\mathrm{H}}(t)^{2}\text{,}%
\end{equation}
that is,%
\begin{equation}
\frac{\hslash^{4}}{16}\frac{4}{\hslash^{2}}\Delta\mathrm{H}^{2}a_{\mathrm{H}%
}(t)^{2}\leq\Delta\mathrm{H}^{2}\Delta\mathrm{\dot{H}}^{2}-\frac{1}%
{4}\left\vert \left\langle \left[  \mathrm{H}\text{, }\mathrm{\dot{H}}\right]
\right\rangle \right\vert ^{2} \label{luo3}%
\end{equation}
Finally, algebraic manipulation of Eq. (\ref{luo3}) yields%
\begin{equation}
a_{\mathrm{H}}(t)^{2}\leq\frac{4}{\hslash^{2}}\Delta\mathrm{\dot{H}}^{2}%
-\frac{1}{\hslash^{2}}\frac{1}{\Delta\mathrm{H}^{2}}\left\vert \left\langle
\left[  \mathrm{H}\text{, }\mathrm{\dot{H}}\right]  \right\rangle \right\vert
^{2}\text{,} \label{king}%
\end{equation}
that is,%
\begin{equation}
a_{\mathrm{H}}(t)^{2}\leq\frac{4}{\hslash^{2}}\Delta\mathrm{\dot{H}}%
^{2}\text{,} \label{luo4}%
\end{equation}
or, equivalently,%
\begin{equation}
\left\vert a_{\mathrm{H}}(t)\right\vert \leq\frac{2}{\hslash}\Delta
\mathrm{\dot{H}}\left(  t\right)  \text{.} \label{luo5}%
\end{equation}
Eq. (\ref{luo5}) ends Pati's main derivation and expresses the fact that the
modulus of the acceleration of the quantum evolution is upper bounded by the
fluctuation in the temporal rate of change of the Hamiltonian of the system
under consideration.

After discussing Pati's derivation, we are ready to discuss the Alsing-Cafaro
derivation which uses a different geometric approach but leads to the same
upper bounds on quantum acceleration.

\subsection{The Alsing-Cafaro derivation}

We begin by pointing out that in the Alsing-Cafaro derivation it is assumed
$\hslash=1$ and the Fubini-Study infinitesimal line element written as
$ds_{\mathrm{FS}}^{2}\overset{\text{def}}{=}\sigma_{\mathrm{H}}^{2}dt^{2}$,
where $\sigma_{\mathrm{H}}^{2}\overset{\text{def}}{=}\left\langle \left(
\Delta\mathrm{H}\right)  ^{2}\right\rangle $ is the variance of the
Hamiltonian operator \textrm{H} with $\Delta\mathrm{H}\overset{\text{def}}%
{=}\mathrm{H}-\left\langle \mathrm{H}\right\rangle $. For clarity, we remark
that while $\left(  \Delta\mathrm{H}\right)  _{\mathrm{Pati}}^{2}%
\overset{\text{def}}{=}\left\langle \mathrm{H}^{2}\right\rangle -\left\langle
\mathrm{H}\right\rangle ^{2}$ is a scalar quantity in Ref. \cite{pati23},
$\left(  \Delta\mathrm{H}\right)  _{\text{\textrm{Alsing}-\textrm{Cafaro}}%
}^{2}\overset{\text{def}}{=}\left(  \mathrm{H}-\left\langle \mathrm{H}%
\right\rangle \right)  ^{2}$ is an operator in Ref. \cite{alsing24}. In this
context, the speed of the quantum evolution in projective Hilbert space is
$v_{\mathrm{H}}\overset{\text{def}}{=}\sigma_{\mathrm{H}}$, while
$a_{\mathrm{H}}\overset{\text{def}}{=}\partial_{t}v_{\mathrm{H}}=\partial
_{t}\sigma_{\mathrm{H}}=\dot{\sigma}_{\mathrm{H}}$ denotes the acceleration of
the quantum evolution. Then, for any finite-dimensional quantum system that
evolves under the time-dependent Hamiltonian \textrm{H}$(t)$, the goal is to
show that the following inequality holds true%
\begin{equation}
\left(  \partial_{t}\sigma_{\mathrm{H}}\right)  ^{2}\leq\left(  \sigma
_{\partial_{t}\mathrm{H}}\right)  ^{2}\text{,} \label{one}%
\end{equation}
that is, $\left\vert a_{\mathrm{H}}\left(  t\right)  \right\vert \leq
\sigma_{\mathrm{\dot{H}}}$, with $\sigma_{\mathrm{\dot{H}}}^{2}=\sigma
_{\partial_{t}\mathrm{H}}^{2}=\left\langle \mathrm{\dot{H}}^{2}\right\rangle
-\left\langle \mathrm{\dot{H}}\right\rangle ^{2}$ denoting the dispersion of
the time-derivative $\mathrm{\dot{H}}$ of the nonstationary Hamiltonian
operator $\mathrm{H}=\mathrm{H}(t)$. Alsing and Cafaro demonstrated that the
inequality in Eq. (\ref{one}) is a consequence of the Robertson uncertainty
relation which, in turn, is also known as a sort of generalized uncertainty
principle in quantum theory since it extends its application to variables that
are not necessarily conjugate. Following Ref. \cite{sakurai85,hall13}, the
Robertson uncertainty relation expresses the fact that any pair of quantum
observables $A$ and $B$ fulfill the inequality%
\begin{equation}
\left\langle \left(  \Delta A\right)  ^{2}\right\rangle \left\langle \left(
\Delta B\right)  ^{2}\right\rangle \geq\frac{1}{4}\left\vert \left\langle
\left[  A\text{, }B\right]  \right\rangle \right\vert ^{2}\text{,}
\label{sakurai}%
\end{equation}
with the expectation values in Eq. (\ref{sakurai}) being evaluated with
respect to a fixed physical state. We start by noting that since $\left(
\partial_{t}\sigma_{\mathrm{H}}\right)  ^{2}=\left(  \partial_{t}\left\langle
\left(  \Delta\mathrm{H}\right)  ^{2}\right\rangle \right)  ^{2}/4\left\langle
\left(  \Delta\mathrm{H}\right)  ^{2}\right\rangle $ and $\left(
\sigma_{\partial_{t}\mathrm{H}}\right)  ^{2}=\left\langle \left(
\Delta\mathrm{\dot{H}}\right)  ^{2}\right\rangle $, Eq. (\ref{one}) can be
rewritten as%
\begin{equation}
\left\langle \left(  \Delta\mathrm{\dot{H}}\right)  ^{2}\right\rangle
\left\langle \left(  \Delta\mathrm{H}\right)  ^{2}\right\rangle \geq\frac
{1}{4}\left[  \partial_{t}\left\langle \left(  \Delta\mathrm{H}\right)
^{2}\right\rangle \right]  ^{2}\text{.} \label{love1}%
\end{equation}
Interestingly, one observes that the term $\left[  \partial_{t}\left\langle
\left(  \Delta\mathrm{H}\right)  ^{2}\right\rangle \right]  ^{2}$ in Eq.
(\ref{love1}) can be suitably recast by means of a quantum anti-commutator,%
\begin{equation}
\left[  \partial_{t}\left\langle \left(  \Delta\mathrm{H}\right)
^{2}\right\rangle \right]  ^{2}=\left[  \left\langle \left(  \Delta
\mathrm{\dot{H}}\right)  \left(  \Delta\mathrm{H}\right)  +\left(
\Delta\mathrm{H}\right)  \left(  \Delta\mathrm{\dot{H}}\right)  \right\rangle
\right]  ^{2}=\left\langle \left\{  \Delta\mathrm{\dot{H}}\text{, }%
\Delta\mathrm{H}\right\}  \right\rangle ^{2}\text{.} \label{love2}%
\end{equation}
Therefore, making a joint use of Eqs. (\ref{love1}) and (\ref{love2}), the
inequality in\ Eq. (\ref{one}) reduces to%
\begin{equation}
\left\langle \left(  \Delta\mathrm{\dot{H}}\right)  ^{2}\right\rangle
\left\langle \left(  \Delta\mathrm{H}\right)  ^{2}\right\rangle \geq\frac
{1}{4}\left\langle \left\{  \Delta\mathrm{\dot{H}}\text{, }\Delta
\mathrm{H}\right\}  \right\rangle ^{2}\text{.} \label{b}%
\end{equation}
From Eq. (\ref{b}), one observes that $\left\langle \left(  \Delta
\mathrm{\dot{H}}\right)  ^{2}\right\rangle \left\langle \left(  \Delta
\mathrm{H}\right)  ^{2}\right\rangle =\left\langle \Delta\mathrm{\dot{H}%
}\left\vert \Delta\mathrm{\dot{H}}\right.  \right\rangle \left\langle
\Delta\mathrm{H}\left\vert \Delta\mathrm{H}\right.  \right\rangle $. For
clarity, we point out that in our notation, we have $\left\langle \left(
\Delta\mathrm{\dot{H}}\right)  ^{2}\right\rangle =\left\langle \left(
\Delta\mathrm{\dot{H}}\right)  \left(  \Delta\mathrm{\dot{H}}\right)
\right\rangle =\left\langle \psi\left\vert \left(  \Delta\mathrm{\dot{H}%
}\right)  \left(  \Delta\mathrm{\dot{H}}\right)  \right\vert \psi\right\rangle
=\left(  \left\langle \psi\right\vert \Delta\mathrm{\dot{H}}\right)  \left(
\Delta\mathrm{\dot{H}}\left\vert \psi\right\rangle \right)  =\left\langle
\Delta\mathrm{\dot{H}}\left\vert \Delta\mathrm{\dot{H}}\right.  \right\rangle
$. Clearly, $\left\vert \psi\right\rangle $ is the state vector of the system,
$\Delta\mathrm{\dot{H}}$ is an operator, and $\Delta\mathrm{\dot{H}}\left\vert
\psi\right\rangle $ is a vector. Thus,\textbf{ }$\left\langle \Delta
\mathrm{\dot{H}}\left\vert \Delta\mathrm{\dot{H}}\right.  \right\rangle $ is a
shorthand for the overlap of the state $\Delta\mathrm{\dot{H}}\left\vert
\psi\right\rangle $ with itself. Then, exploiting the Schwarz inequality
\cite{sakurai85}, one gets%
\begin{equation}
\left\langle \Delta\mathrm{\dot{H}}\left\vert \Delta\mathrm{\dot{H}}\right.
\right\rangle \left\langle \Delta\mathrm{H}\left\vert \Delta\mathrm{H}\right.
\right\rangle \geq\left\vert \left\langle \left(  \Delta\mathrm{H}\right)
\left(  \Delta\mathrm{\dot{H}}\right)  \right\rangle \right\vert ^{2}\text{,}%
\end{equation}
that is,%
\begin{equation}
\left\langle \left(  \Delta\mathrm{\dot{H}}\right)  ^{2}\right\rangle
\left\langle \left(  \Delta\mathrm{H}\right)  ^{2}\right\rangle \geq\left\vert
\left\langle \left(  \Delta\mathrm{H}\right)  \left(  \Delta\mathrm{\dot{H}%
}\right)  \right\rangle \right\vert ^{2}\text{.} \label{b4}%
\end{equation}
From Eqs. (\ref{b}) and (\ref{b4}), one can arrive at the conclusion that the
inequality in Eq. (\ref{b}) can be demonstrated if one is capable of verifying
that
\begin{equation}
\left\vert \left\langle \left(  \Delta\mathrm{H}\right)  \left(
\Delta\mathrm{\dot{H}}\right)  \right\rangle \right\vert ^{2}\geq\frac{1}%
{4}\left\langle \left\{  \Delta\mathrm{\dot{H}}\text{, }\Delta\mathrm{H}%
\right\}  \right\rangle ^{2}\text{.} \label{c}%
\end{equation}
Therefore, let us aim to prove the inequality in Eq. (\ref{c}). Observe that
$2\left(  \Delta\mathrm{H}\right)  \left(  \Delta\mathrm{\dot{H}}\right)  $
can be recast as a sum of a commutator and an anti-commutator,%
\begin{equation}
2\left(  \Delta\mathrm{H}\right)  \left(  \Delta\mathrm{\dot{H}}\right)
=\left[  \Delta\mathrm{H}\text{, }\Delta\mathrm{\dot{H}}\right]  +\left\{
\Delta\mathrm{H}\text{, }\Delta\mathrm{\dot{H}}\right\}  =\left[
\Delta\mathrm{H}\text{, }\Delta\mathrm{\dot{H}}\right]  +\left\{
\Delta\mathrm{\dot{H}}\text{, }\Delta\mathrm{H}\right\}  \text{.} \label{d1}%
\end{equation}
Since $\left[  \Delta\mathrm{H}\text{, }\Delta\mathrm{\dot{H}}\right]  $
represents an anti-Hermitian operator, $\ \left\langle \left[  \Delta
\mathrm{H}\text{, }\Delta\mathrm{\dot{H}}\right]  \right\rangle $ is purely
imaginary. Furthermore, since $\left\{  \Delta\mathrm{\dot{H}}\text{, }%
\Delta\mathrm{H}\right\}  $ denotes a Hermitian operator, its expectation
value $\left\langle \left\{  \Delta\mathrm{\dot{H}}\text{, }\Delta
\mathrm{H}\right\}  \right\rangle $ is real. Therefore, using these two
properties, from Eq. (\ref{d1}) one arrives at the relation%
\begin{equation}
4\left\vert \left\langle \left(  \Delta\mathrm{H}\right)  \left(
\Delta\mathrm{\dot{H}}\right)  \right\rangle \right\vert ^{2}=\left\vert
\left\langle \left[  \Delta\mathrm{H}\text{, }\Delta\mathrm{\dot{H}}\right]
\right\rangle \right\vert ^{2}+\left\vert \left\langle \left\{  \Delta
\mathrm{\dot{H}}\text{, }\Delta\mathrm{H}\right\}  \right\rangle \right\vert
^{2}\text{,}%
\end{equation}
that is,%
\begin{equation}
\left\vert \left\langle \left(  \Delta\mathrm{H}\right)  \left(
\Delta\mathrm{\dot{H}}\right)  \right\rangle \right\vert ^{2}=\frac{\left\vert
\left\langle \left[  \Delta\mathrm{H}\text{, }\Delta\mathrm{\dot{H}}\right]
\right\rangle \right\vert ^{2}+\left\vert \left\langle \left\{  \Delta
\mathrm{\dot{H}}\text{, }\Delta\mathrm{H}\right\}  \right\rangle \right\vert
^{2}}{4}\geq\frac{1}{4}\left\vert \left\langle \left\{  \Delta\mathrm{\dot{H}%
}\text{, }\Delta\mathrm{H}\right\}  \right\rangle \right\vert ^{2}\text{.}
\label{d}%
\end{equation}
From Eq. (\ref{d}), one arrives at the conclusion that the inequality in Eq.
(\ref{c}) is fulfilled and, thus, our inequality in Eq. (\ref{b}) is also
demonstrated. Finally, since the inequalities in Eqs. (\ref{one}) and
(\ref{b}) are equivalent, one obtains
\begin{equation}
\left\vert a_{\mathrm{H}}\left(  t\right)  \right\vert \leq\sigma
_{\mathrm{\dot{H}}}\left(  t\right)  \text{.} \label{cicca}%
\end{equation}
Eq. (\ref{cicca}) ends the Alsing-Cafaro main derivation and expresses the
fact that the modulus of the acceleration $\left\vert a_{\mathrm{H}}\left(
t\right)  \right\vert $ of the quantum evolution is upper bounded by the
fluctuation $\sigma_{\mathrm{\dot{H}}}$ in the temporal rate of change of the
Hamiltonian of the system under consideration. Inspecting Eqs. (\ref{luo5})
and (\ref{cicca}), we conclude that both Pati's and the Alsing-Cafaro
derivations yield the same quantum acceleration limit once some notational
differences and simplifying working conditions are spotted. In particular,
while $\hslash$ is set equal to one in the Alsing-Cafaro approach, Pati keeps
$\hslash\neq1$. Furthermore, the Fubini-Study line element (squared) in Pati's
derivation is defined as four times the Fubini-Study line element (squared)
employed in the Alsing-Cafaro analysis. Finally, while Alsing and Cafaro use
the Robertson uncertainty relation in their analysis, Pati employes the
Robertson-Schr\"{o}dinger uncertainty inequality in his derivation. A visual
summary of the main differences between Pati's and the Alsing-Cafaro
derivations of a quantum acceleration limit appears in Table
I.\begin{table}[t]
\centering
\begin{tabular}
[c]{c|c|c|c}\hline\hline
\textbf{Derivations} & \textbf{Reduced Planck constant} & \textbf{Fubini-Study
line element} & \textbf{Uncertainty relation}\\\hline
Pati & $\hslash\neq1$ & $ds_{\mathrm{FS}}^{2}\overset{\text{def}}{=}\frac
{4}{\hslash^{2}}\Delta\mathrm{H}\left(  t\right)  ^{2}dt^{2}$ &
Robertson-Schr\"{o}dinger\\\hline
Alsing-Cafaro & $\hslash=1$ & $ds_{\mathrm{FS}}^{2}\overset{\text{def}}%
{=}\Delta\mathrm{H}\left(  t\right)  ^{2}dt^{2}$ & Robertson\\\hline
\end{tabular}
\caption{Schematic summary of the main differences between Pati's and the
Alsing-Cafaro derivations leading, essentially, to the very same quantum
acceleration limit. Note that $\Delta\mathrm{H}\left(  t\right)  ^{2}%
\overset{\text{def}}{=}\left\langle \mathrm{H}\left(  t\right)  ^{2}%
\right\rangle -\left\langle \mathrm{H}\left(  t\right)  \right\rangle
^{2}=\sigma_{\mathrm{H}}^{2}\left(  t\right)  $, with the averages taken with
respect to a (normalized) state $\left\vert \psi\left(  t\right)
\right\rangle $.}%
\end{table}

We are now ready to discuss quantum acceleration limits for the evolution of
two-level quantum systems.

\section{Quantum acceleration limit for qubits}

The quantum acceleration limits discussed in the previous section applies to
systems characterized by an arbitrary finite-dimensional projective Hilbert
space. In this section, instead, we focus on a geometric description of the
quantum evolution of two-level quantum systems on a Bloch sphere under general
time-dependent Hamiltonians. In particular, we find the most general necessary
and sufficient conditions needed to reach the maximal upper bounds on the
acceleration of the quantum evolution. These conditions are expressed
explicitly in terms of two three-dimensional real vectors, the Bloch vector
$\mathbf{a=a}\left(  t\right)  $ that corresponds to the evolving quantum
state $\left\vert \psi\left(  t\right)  \right\rangle $ with $\rho\left(
t\right)  \overset{\text{def}}{=}\left\vert \psi\left(  t\right)
\right\rangle \left\langle \psi\left(  t\right)  \right\vert =(1/2)\left(
\mathbf{1+a\cdot\boldsymbol{\sigma}}\right)  $ and the magnetic field vector
$\mathbf{h=h}\left(  t\right)  $ that specifies the (traceless) Hermitian
Hamiltonian \textrm{H}$\left(  t\right)  \overset{\text{def}}{=}$
$\mathbf{h}\left(  t\right)  \cdot\mathbf{\boldsymbol{\sigma}}$ of the system.

We begin with a formal discussion here and place examples in the next section.
In the Alsing-Cafaro approach, with $\hslash=1$ and $\left(  ds_{\mathrm{FS}%
}^{2}\right)  _{\text{\textrm{Alsing}-}\mathrm{Cafaro}}=(1/4)\left(
ds_{\mathrm{FS}}^{2}\right)  _{\mathrm{Pati}}$, the inequality in Eq.
(\ref{king}) becomes
\begin{equation}
a_{\mathrm{H}}\left(  t\right)  ^{2}\leq\sigma_{\mathrm{\dot{H}}}\left(
t\right)  ^{2}-\frac{1}{4}\frac{1}{\sigma_{\mathrm{H}}\left(  t\right)  ^{2}%
}\left\vert \left\langle \left[  \mathrm{H}\left(  t\right)  \text{,
}\mathrm{\dot{H}}\left(  t\right)  \right]  \right\rangle \right\vert
^{2}\text{.} \label{kingineq}%
\end{equation}
The goal is to express the inequality in Eq. (\ref{kingineq}), valid for
arbitrary finite-dimensional quantum systems in a pure state, in terms of an
algebraic inequality that describes geometric constraints on the vectors
$\mathbf{a=a}\left(  t\right)  $ and $\mathbf{h=h}\left(  t\right)  $. Note
that $a_{\mathrm{H}}\left(  t\right)  ^{2}=\left[  \partial_{t}v_{\mathrm{H}%
}\left(  t\right)  \right]  ^{2}=\left[  \partial_{t}\sigma_{\mathrm{H}%
}\left(  t\right)  \right]  ^{2}$, with $\sigma_{\mathrm{H}}^{2}%
\overset{\text{def}}{=}\left\langle \left(  \mathrm{H}-\left\langle
\mathrm{H}\right\rangle \right)  ^{2}\right\rangle =\left\langle
\mathrm{H}^{2}\right\rangle -\left\langle \mathrm{H}\right\rangle
^{2}=\mathrm{tr}\left(  \rho\mathrm{H}^{2}\right)  -\left[  \mathrm{tr}\left(
\rho\mathrm{H}\right)  \right]  ^{2}$. Using the fact that $\rho\left(
t\right)  \overset{\text{def}}{=}(1/2)\left(  \mathbf{1+a\cdot
\boldsymbol{\sigma}}\right)  $ with $\mathbf{1}$ being the identity operator
and \textrm{H}$\left(  t\right)  \overset{\text{def}}{=}$ $\mathbf{h}%
\cdot\mathbf{\boldsymbol{\sigma}}$, a straightforward calculation (see
Appendix B for details) yields%
\begin{equation}
\sigma_{\mathrm{H}}^{2}=\mathbf{h}^{2}-\left(  \mathbf{a}\cdot\mathbf{h}%
\right)  ^{2}\text{.} \label{sigma1}%
\end{equation}
From Eq. (\ref{sigma1}), we can calculate $a_{\mathrm{H}}\left(  t\right)
=\partial_{t}v_{\mathrm{H}}\left(  t\right)  =\partial_{t}\sigma_{\mathrm{H}%
}\left(  t\right)  $. Recalling that since $\mathbf{\dot{a}}\cdot\mathbf{h=0}$
since $\mathbf{\dot{a}}$ satisfies the equation $\mathbf{\dot{a}=2h\times a}$
(for details, see Appendix A in Ref. \cite{alsing24}), a simple calculation
leads to%
\begin{equation}
a_{\mathrm{H}}=\frac{\mathbf{h\cdot\dot{h}-}\left(  \mathbf{a}\cdot
\mathbf{h}\right)  \left(  \mathbf{a}\cdot\mathbf{\dot{h}}\right)  }%
{\sqrt{\mathbf{h}^{2}-\left(  \mathbf{a}\cdot\mathbf{h}\right)  ^{2}}}\text{.}
\label{sigma}%
\end{equation}
Next, we need to find an expression for $\sigma_{\mathrm{\dot{H}}}^{2}%
\overset{\text{def}}{=}\left\langle \left(  \mathrm{\dot{H}}-\left\langle
\mathrm{\dot{H}}\right\rangle \right)  ^{2}\right\rangle =\left\langle
\mathrm{\dot{H}}^{2}\right\rangle -\left\langle \mathrm{\dot{H}}\right\rangle
^{2}=\mathrm{tr}(\rho\mathrm{\dot{H}}^{2})-\left[  \mathrm{tr}\left(
\rho\mathrm{\dot{H}}\right)  \right]  ^{2}$. Again, making use of the fact
that $\rho\left(  t\right)  \overset{\text{def}}{=}(1/2)\left(
\mathbf{1+a\cdot\boldsymbol{\sigma}}\right)  $ and \textrm{H}$\left(
t\right)  \overset{\text{def}}{=}$ $\mathbf{h}\cdot\mathbf{\boldsymbol{\sigma
}}$, we obtain%
\begin{equation}
\sigma_{\mathrm{\dot{H}}}^{2}=\mathbf{\dot{h}}^{2}-\left(  \mathbf{a}%
\cdot\mathbf{\dot{h}}\right)  ^{2}\text{.} \label{sigma2}%
\end{equation}
We remark that the derivation of Eq. (\ref{sigma2}) is formally identical to
the derivation of Eq. (\ref{sigma1}). For this reason, we refer to Appendix B
for technical details. Lastly, we need an expression for $\left\vert
\left\langle \left[  \mathrm{H}\left(  t\right)  \text{, }\mathrm{\dot{H}%
}\left(  t\right)  \right]  \right\rangle \right\vert ^{2}$ in Eq.
(\ref{kingineq}). From \textrm{H}$\overset{\text{def}}{=}$ $\mathbf{h}%
\cdot\mathbf{\boldsymbol{\sigma}}$ and \textrm{\.{H}}$\overset{\text{def}}{=}$
$\mathbf{\dot{h}}\cdot\mathbf{\boldsymbol{\sigma}}$, a straightforward but
tedious calculation yields (for details, see Appendix B)%
\begin{equation}
\left\vert \left\langle \left[  \mathrm{H}\text{, }\mathrm{\dot{H}}\right]
\right\rangle \right\vert ^{2}=\left[  \left(  \mathbf{a\times h}\right)
\cdot\mathbf{\dot{h}-}\left(  \mathbf{a\times\dot{h}}\right)  \cdot
\mathbf{h}\right]  ^{2}\text{.} \label{sigma3}%
\end{equation}
Observe that, writing $\mathbf{h}\left(  t\right)  =h(t)\hat{h}(t)$, we have
$\mathbf{\dot{h}}\left(  t\right)  =\dot{h}(t)\hat{h}(t)+h(t)\partial_{t}%
\hat{h}(t)$. Therefore, we conclude that $\left\vert \left\langle \left[
\mathrm{H}\text{, }\mathrm{\dot{H}}\right]  \right\rangle \right\vert ^{2}$ in
Eq. (\ref{sigma3}) vanishes when the magnetic field $\mathbf{h}\left(
t\right)  $ does not change in direction, that is $\partial_{t}\hat
{h}(t)=\mathbf{0}$ (i.e., $\mathbf{h}\left(  t\right)  $ and $\mathbf{\dot{h}%
}\left(  t\right)  $ are collinear). We are now in a position of being able to
express the inequality in Eq. (\ref{kingineq}) in terms of $\mathbf{a}$ and
$\mathbf{h}$. Indeed, using Eqs. (\ref{sigma1}), (\ref{sigma}), (\ref{sigma2}%
), and (\ref{sigma3}), Eq. (\ref{kingineq}) reduces to%
\begin{equation}
\frac{\left[  \mathbf{h\cdot\dot{h}-}\left(  \mathbf{a}\cdot\mathbf{h}\right)
\left(  \mathbf{a}\cdot\mathbf{\dot{h}}\right)  \right]  ^{2}}{\mathbf{h}%
^{2}-\left(  \mathbf{a}\cdot\mathbf{h}\right)  ^{2}}\leq\left[  \mathbf{\dot
{h}}^{2}-\left(  \mathbf{a}\cdot\mathbf{\dot{h}}\right)  ^{2}\right]
-\frac{1}{4}\frac{\left[  \left(  \mathbf{a\times h}\right)  \cdot
\mathbf{\dot{h}-}\left(  \mathbf{a\times\dot{h}}\right)  \cdot\mathbf{h}%
\right]  ^{2}}{\mathbf{h}^{2}-\left(  \mathbf{a}\cdot\mathbf{h}\right)  ^{2}%
}\text{,} \label{king2}%
\end{equation}
or, more loosely,%
\begin{equation}
\frac{\left[  \mathbf{h\cdot\dot{h}-}\left(  \mathbf{a}\cdot\mathbf{h}\right)
\left(  \mathbf{a}\cdot\mathbf{\dot{h}}\right)  \right]  ^{2}}{\mathbf{h}%
^{2}-\left(  \mathbf{a}\cdot\mathbf{h}\right)  ^{2}}\leq\mathbf{\dot{h}}%
^{2}-\left(  \mathbf{a}\cdot\mathbf{\dot{h}}\right)  ^{2}\text{.}
\label{king3}%
\end{equation}
We should be able to explicitly check the inequality in Eq. (\ref{king3}) by
simply using the fact that $\mathbf{a}\cdot\mathbf{a=}1$ and $\mathbf{\dot
{a}=2h\times a}$.

\emph{Case}:\emph{ }$\mathbf{a}\cdot\mathbf{h=}0$. If we assume $\mathbf{a}%
\cdot\mathbf{h=}0$, then $\partial_{t}(\mathbf{a}\cdot\mathbf{h})=0$.
Therefore, $\mathbf{\dot{a}}\cdot\mathbf{h+a}\cdot\mathbf{\dot{h}=}0$.
However, since $\mathbf{\dot{a}}\cdot\mathbf{h=0}$ because $\mathbf{\dot{a}}$
satisfies the equation $\mathbf{\dot{a}=2h\times a}$, we have $\mathbf{a}%
\cdot\mathbf{\dot{h}=}0$ when $\mathbf{a}\cdot\mathbf{h=}0$ is assumed. In
this case, the inequality in Eq. (\ref{king3}) reduces to $\left(
\mathbf{h\cdot\dot{h}}\right)  ^{2}\leq\mathbf{\dot{h}}^{2}\mathbf{h}^{2}$.
The inequality, in turn, is obviously correct. The take-home message when
$\mathbf{a}\cdot\mathbf{h=}0$ is the following. When we assume $\mathbf{a}%
\cdot\mathbf{h=}0$, the set $\left\{  \mathbf{a}\text{, }\mathbf{\dot{a}%
}\text{, }\mathbf{h}\right\}  $ is a set of orthogonal vectors. In particular,
in this case, $\mathbf{h}$ and $\mathbf{\dot{h}}$ do not need to be collinear.
We only have that $\mathbf{\dot{h}}$ is orthogonal to $\mathbf{a}$ and belongs
to the plane spanned by $\left\{  \mathbf{\dot{a}}\text{, }\mathbf{h}\right\}
$. The orthogonality between $\mathbf{\dot{h}}$ and $\mathbf{a}$ is a
consequence of the assumption $\mathbf{a}\cdot\mathbf{h=}0$. Also, $\left(
a_{\mathrm{H}}^{2}\right)  _{\max}=\mathbf{\dot{h}}^{2}$ is achieved only when
$\mathbf{h}$ and $\mathbf{\dot{h}}$ are collinear (i.e., the magnetic field
does not change in direction). Finally, the collinearity of $\mathbf{h}$ and
$\mathbf{\dot{h}}$ also implies the vanishing of the term $\left\vert
\left\langle \left[  \mathrm{H}\text{, }\mathrm{\dot{H}}\right]  \right\rangle
\right\vert ^{2}$ in Eq. (\ref{king2}).

\emph{Case}: $\mathbf{a}\cdot\mathbf{h\neq}0$. Note that the inequality in Eq.
(\ref{king3}) can be recast as
\begin{equation}
\left[  \mathbf{h\cdot\dot{h}-}\left(  \mathbf{a}\cdot\mathbf{h}\right)
\left(  \mathbf{a}\cdot\mathbf{\dot{h}}\right)  \right]  ^{2}\leq\left[
\mathbf{\dot{h}}^{2}-\left(  \mathbf{a}\cdot\mathbf{\dot{h}}\right)
^{2}\right]  \left[  \mathbf{h}^{2}-\left(  \mathbf{a}\cdot\mathbf{h}\right)
^{2}\right]  \text{.} \label{h1}%
\end{equation}
A simple calculation shows that for collinear vectors $\mathbf{h}$ and
$\mathbf{\dot{h}}$, the inequality in Eq. (\ref{h1}) is saturated. Therefore,
the condition $\mathbf{h\times\dot{h}=0}$ is a sufficient condition for
obtaining saturation. However, is it necessary? We explore this question now.
\ Let us decompose the vectors $\mathbf{h}$ and $\mathbf{\dot{h}}$ as%
\begin{equation}
\mathbf{h}\overset{\text{def}}{\mathbf{=}}\left(  \mathbf{h\cdot a}\right)
\mathbf{a+}\left[  \mathbf{h-}\left(  \mathbf{h\cdot a}\right)  \mathbf{a}%
\right]  \text{, and }\mathbf{\dot{h}}\overset{\text{def}}{\mathbf{=}}\left(
\mathbf{\dot{h}\cdot a}\right)  \mathbf{a+}\left[  \mathbf{\dot{h}-}\left(
\mathbf{\dot{h}\cdot a}\right)  \mathbf{a}\right]  \text{,} \label{h2}%
\end{equation}
respectively. Then, consider
\begin{equation}
\mathbf{h}\overset{\text{def}}{=}\mathbf{h}_{\Vert}\mathbf{+h}_{\perp}\text{,
with }\mathbf{h}_{\Vert}\overset{\text{def}}{=}\left(  \mathbf{h\cdot
a}\right)  \mathbf{a}\text{ and }\mathbf{h}_{\perp}\overset{\text{def}}%
{=}\mathbf{h-}\left(  \mathbf{h\cdot a}\right)  \mathbf{a}\text{,} \label{h3}%
\end{equation}
and%
\begin{equation}
\mathbf{\dot{h}}\overset{\text{def}}{=}\mathbf{\dot{h}}_{\Vert}\mathbf{+\dot
{h}}_{\perp}\text{, with }\mathbf{\dot{h}}_{\Vert}\overset{\text{def}}%
{=}\left(  \mathbf{\dot{h}\cdot a}\right)  \mathbf{a}\text{ and }%
\mathbf{\dot{h}}_{\perp}\overset{\text{def}}{=}\mathbf{\dot{h}-}\left(
\mathbf{\dot{h}\cdot a}\right)  \mathbf{a}\text{.} \label{h4}%
\end{equation}
Using Eqs. (\ref{h2}), (\ref{h3}), and (\ref{h4}), the inequality in Eq.
(\ref{h1}) becomes%
\begin{equation}
\left[  \left(  h_{\Vert}\hat{a}+\mathbf{h}_{\perp}\right)  \cdot\left(
\dot{h}_{\Vert}\hat{a}+\mathbf{\dot{h}}_{\perp}\right)  -h_{\Vert}\dot
{h}_{\Vert}\right]  ^{2}\leq\mathbf{\dot{h}}_{\perp}^{2}\mathbf{h}_{\perp}%
^{2}\text{.} \label{h4b}%
\end{equation}
Since $\hat{a}\cdot\hat{a}=1$, $\mathbf{h}_{\perp}\cdot\mathbf{a=}0$, and
$\mathbf{\dot{h}}_{\perp}\cdot\mathbf{a=}0$, the left-hand-side of the
inequality in Eq. (\ref{h4b}) reduces to $\left(  \mathbf{h}_{\perp}%
\cdot\mathbf{\dot{h}}_{\perp}\right)  ^{2}$. Therefore, Eq. (\ref{h4b}) yields%
\begin{equation}
\left(  \mathbf{h}_{\perp}\cdot\mathbf{\dot{h}}_{\perp}\right)  ^{2}%
\leq\mathbf{\dot{h}}_{\perp}^{2}\mathbf{h}_{\perp}^{2}\text{.} \label{h5}%
\end{equation}
The inequality in Eq. (\ref{h5}) is clearly correct. Thus, the thesis follows
since we were able to explicitly check the correctness of the inequality in
Eq. (\ref{king3}). As a side remark, we point out that the abstract proofs
presented in the previous section, concerning the quantum acceleration upper
bounds in arbitrary finite-dimensional Hilbert spaces, seem to be more
straightforward than the explicit \textquotedblleft
vectorial\textquotedblright\ proofs restricted to two-level quantum systems
that we have presented in this section. Interestingly, note that the
inequality in Eq. (\ref{h5}) saturates when $\mathbf{h}_{\perp}$ and
$\mathbf{\dot{h}}_{\perp}$ are collinear, that is, when $\mathbf{h}_{\perp
}\times$ $\mathbf{\dot{h}}_{\perp}=\mathbf{0}$. Therefore, we arrive at the
conclusion that the necessary condition for saturation is $\mathbf{h}_{\perp
}\times$ $\mathbf{\dot{h}}_{\perp}=\mathbf{0}$. As a side remark, we also
stress that one can explicitly verify that $\mathbf{h}\times$ $\mathbf{\dot
{h}}=\mathbf{0}$ implies $\mathbf{h}_{\perp}\times$ $\mathbf{\dot{h}}_{\perp
}=\mathbf{0}$ using Eqs. (\ref{h3}) and (\ref{h4}). Moreover, observe that%
\begin{align}
\mathbf{0}  &  =\mathbf{h}_{\perp}\times\mathbf{\dot{h}}_{\perp}\nonumber\\
&  =\left[  h_{\bot}(t)\hat{h}_{\bot}(t)\right]  \times\left\{  \dot{h}_{\bot
}(t)\hat{h}_{\bot}(t)+h_{\bot}(t)\left[  \partial_{t}\hat{h}_{\bot}(t)\right]
\right\} \nonumber\\
&  =h_{\bot}^{2}(t)\left[  \hat{h}_{\bot}(t)\times\partial_{t}\hat{h}_{\bot
}(t)\right]  \text{,}%
\end{align}
implies $\partial_{t}\hat{h}_{\bot}(t)=\vec{0}$ (that is, $\hat{h}_{\bot}$
does not change in time) since we assume $h_{\bot}(t)\neq0$ and $\hat{h}%
_{\bot}(t)\cdot\hat{h}_{\bot}(t)=1$ signifies that $\hat{h}_{\bot}(t)$ and
$\partial_{t}\hat{h}_{\bot}(t)$ are orthogonal vectors. In summary, the
saturation of the inequality in Eq. (\ref{h5}) occurs when $\hat{h}_{\bot}(t)$
is constant in time. An alternative way to arrive at this condition for the
saturation of the inequality in Eq. (\ref{h5}) is as follows. First, note that
the inequality $a_{\mathrm{H}}^{2}\leq\sigma_{\mathrm{\dot{H}}}^{2}$ can also
be recast as,%
\begin{equation}
a_{\mathrm{H}}^{2}=\frac{\left(  \mathbf{h}_{\perp}\cdot\mathbf{\dot{h}%
}_{\perp}\right)  ^{2}}{\mathbf{h}_{\perp}^{2}}\leq\mathbf{\dot{h}}_{\perp
}^{2}=\sigma_{\mathrm{\dot{H}}}^{2}\text{.}%
\end{equation}
Then, setting $\mathbf{h}_{\perp}\left(  t\right)  =h_{\bot}(t)\hat{h}_{\bot
}(t)$ and $\mathbf{\dot{h}}_{\perp}\left(  t\right)  =\dot{h}_{\bot}(t)\hat
{h}_{\bot}(t)+h_{\bot}(t)\left[  \partial_{t}\hat{h}_{\bot}(t)\right]  $, we
have%
\begin{align}
\left(  \mathbf{h}_{\perp}\cdot\mathbf{\dot{h}}_{\perp}\right)  ^{2}  &
=\left(  h_{\bot}(t)\hat{h}_{\bot}(t)\cdot\left\{  \dot{h}_{\bot}(t)\hat
{h}_{\bot}(t)+h_{\bot}(t)\left[  \partial_{t}\hat{h}_{\bot}(t)\right]
\right\}  \right)  ^{2}\nonumber\\
&  =\left(  h_{\bot}(t)\dot{h}_{\bot}(t)+h_{\bot}^{2}(t)\left[  \hat{h}_{\bot
}(t)\cdot\partial_{t}\hat{h}_{\bot}(t)\right]  \right)  ^{2}\nonumber\\
&  =h_{\bot}^{2}(t)\dot{h}_{\bot}^{2}(t) \label{stand1}%
\end{align}
and,%
\begin{equation}
\mathbf{h}_{\perp}^{2}\mathbf{\dot{h}}_{\perp}^{2}=h_{\bot}^{2}(t)\left\{
\dot{h}_{\bot}^{2}(t)+h_{\bot}^{2}(t)\left[  \partial_{t}\hat{h}_{\bot
}(t)\cdot\partial_{t}\hat{h}_{\bot}(t)\right]  \right\}  \label{stand2}%
\end{equation}
Then, from Eqs. (\ref{stand1}) and (\ref{stand2}), we get that $\left(
\mathbf{h}_{\perp}\cdot\mathbf{\dot{h}}_{\perp}\right)  ^{2}=\mathbf{h}%
_{\perp}^{2}\mathbf{\dot{h}}_{\perp}^{2}$ if and only if%
\begin{equation}
\partial_{t}\hat{h}_{\bot}(t)\cdot\partial_{t}\hat{h}_{\bot}(t)=0\text{,}%
\end{equation}
that is, iff $\partial_{t}\hat{h}_{\bot}(t)=\vec{0}$ (i.e., iff $\hat{h}%
_{\bot}$ does not change in time). In Table II, we summarize our main findings
by specifying when the acceleration limit in the qubit case is saturated
(i.e., $a_{\mathrm{H}}^{2}=\left(  a_{\mathrm{H}}^{2}\right)  _{\max}$) and,
in addition, when the saturation limit $\left(  a_{\mathrm{H}}^{2}\right)
_{\max}$ achieves its maximum value $\mathcal{A}_{\max}$.

We are now ready for presenting simple illustrative examples relative to the
saturation of the quantum acceleration limit for single-qubit evolutions in
geometric terms.

\begin{table}[t]
\centering
\begin{tabular}
[c]{c|c|c}\hline\hline
\textbf{Geometric conditions} & \textbf{Inequality saturated}, $a_{\mathrm{H}%
}=\left(  a_{\mathrm{H}}\right)  _{\max}$ & \textbf{Maximum} $\left(
a_{\mathrm{H}}\right)  _{\max}=\mathcal{A}_{\max}$\\\hline
$\mathbf{a\cdot h=}0$, and $\mathbf{h\times\dot{h}\neq0}$ & no & no\\\hline
$\mathbf{a\cdot h=}0$, and $\mathbf{h\times\dot{h}=0}$ & yes & yes\\\hline
$\mathbf{a\cdot h\neq}0$, and $\mathbf{h}_{\perp}\mathbf{\times\dot{h}}%
_{\perp}\mathbf{\neq0}$ & no & no\\\hline
$\mathbf{a\cdot h\neq}0$, and $\mathbf{h}_{\perp}\mathbf{\times\dot{h}}%
_{\perp}\mathbf{=0}$ & yes & no\\\hline
\end{tabular}
\caption{Tabular summary specifying the fact that the acceleration limit in
the qubit case is saturated when $\mathbf{h}_{\perp}$ and $\mathbf{\dot{h}%
}_{\perp}$ are collinear (i.e., when $\mathbf{h}_{\perp}$ with $\mathbf{h}%
_{\perp}\mathbf{\times\dot{h}}_{\perp}\mathbf{=0}$ does not change in
direction). Moreover, when the saturation limit $\left(  a_{\mathrm{H}%
}\right)  _{\max}$ is achieved, the maximum value $\mathcal{A}_{\max}$ of this
saturation limit is obtained when $\mathbf{h}_{\parallel}\overset{\text{def}%
}{=}\left(  \mathbf{h\cdot a}\right)  \mathbf{a}$ vanishes (i.e., when
$\mathbf{a\cdot h=}0$ since $\mathbf{a\neq0}$ with $\mathbf{a\cdot a=}1$).}%
\end{table}

\section{Illustrative examples}

In this section, we present three illustrative examples. In the first example,
the acceleration limit is not saturated (i.e., there is a strict inequality
$a_{\mathrm{H}}<\left(  a_{\mathrm{H}}\right)  _{\max}$). In the second
example, the acceleration limit is saturated, but the saturation value
$\left(  a_{\mathrm{H}}\right)  _{\max}$ does not reach its maximum value
$\mathcal{A}_{\max}$ (i.e., $a_{\mathrm{H}}=\left(  a_{\mathrm{H}}\right)
_{\max}<\mathcal{A}_{\max}$). Finally, in the third example we have that the
acceleration limit is saturated and, in addition, the saturation value reaches
its maximum value (i.e., $a_{\mathrm{H}}=\left(  a_{\mathrm{H}}\right)
_{\max}=\mathcal{A}_{\max}$).

\subsection{No saturation, $a_{\mathrm{H}}<\left(  a_{\mathrm{H}}\right)
_{\max}$}

To present this first example, we exploit the so-called time-dependent
Hamiltonians with $100\%$ evolution speed efficiency approach developed by
Uzdin and collaborators in Ref. \cite{uzdin12}. Consider a normalized quantum
state defined as $\left\vert \psi\left(  t\right)  \right\rangle
\overset{\text{def}}{=}\cos\left(  \omega_{0}t\right)  \left\vert
0\right\rangle +e^{i\nu_{0}t}\sin\left(  \omega_{0}t\right)  \left\vert
1\right\rangle $, with $\nu_{0}$ and $\omega_{0}$ in $%
\mathbb{R}
$. Recall that any state $\left\vert \Psi\left(  t\right)  \right\rangle $ on
the Bloch sphere can be recast as $\left\vert \Psi\left(  t\right)
\right\rangle =\left\vert \Psi\left(  \theta\left(  t\right)  \text{, }%
\varphi\left(  t\right)  \right)  \right\rangle =\cos\left[  \theta\left(
t\right)  /2\right]  \left\vert 0\right\rangle +e^{i\varphi\left(  t\right)
}\sin\left[  \theta\left(  t\right)  /2\right]  \left\vert 1\right\rangle $,
with $\theta\left(  t\right)  $ and $\varphi\left(  t\right)  $ being the
polar and azimuthal angles, respectively. Recasting $\left\vert \psi\left(
t\right)  \right\rangle $ as $\left\vert \Psi\left(  t\right)  \right\rangle
$, we get $\omega_{0}=\dot{\theta}/2$ and $\nu_{0}=\dot{\varphi}$. \ We
further note that since $\left\langle \psi\left(  t\right)  \left\vert
\dot{\psi}\left(  t\right)  \right.  \right\rangle =i\nu_{0}\sin^{2}\left(
\omega_{0}t\right)  \neq0$, $\left\vert \psi\left(  t\right)  \right\rangle $
is not parallel transported. From $\left\vert \psi\left(  t\right)
\right\rangle $, we construct the state $\left\vert m(t)\right\rangle
\overset{\text{def}}{=}e^{-i\phi\left(  t\right)  }\left\vert \psi\left(
t\right)  \right\rangle $ where the phase $\phi\left(  t\right)  $ is chosen
such that $\left\langle m(t)\left\vert \dot{m}(t)\right.  \right\rangle =0$. A
straightforward calculation shows that $\left\langle m(t)\left\vert \dot
{m}(t)\right.  \right\rangle =0$ iff $-i\dot{\phi}+\left\langle \psi\left(
t\right)  \left\vert \dot{\psi}\left(  t\right)  \right.  \right\rangle =0$,
that is iff $\dot{\phi}=-i\left\langle \psi\left(  t\right)  \left\vert
\dot{\psi}\left(  t\right)  \right.  \right\rangle $. Choosing $\phi\left(
0\right)  =0$, the phase $\phi\left(  t\right)  $ becomes
\begin{equation}
\phi\left(  t\right)  =\frac{\nu_{0}}{4\omega_{0}}\left[  2\omega_{0}%
t-\sin\left(  2\omega_{0}t\right)  \right]  \text{.} \label{s5}%
\end{equation}
Then, employing Eq. (\ref{s5}), the parallel transported normalized state
$\left\vert m(t)\right\rangle $ reduces to%
\begin{equation}
\left\vert m(t)\right\rangle =e^{-i\frac{\nu_{0}}{4\omega_{0}}\left[
2\omega_{0}t-\sin\left(  2\omega_{0}t\right)  \right]  }\left[  \cos\left(
\omega_{0}t\right)  \left\vert 0\right\rangle +e^{i\nu_{0}t}\sin\left(
\omega_{0}t\right)  \left\vert 1\right\rangle \right]  \text{.} \label{s5a}%
\end{equation}
Following the formalism introduced in Ref. \cite{uzdin12}, we set the matrix
representation of the Hamiltonian \textrm{H}$(t)\overset{\text{def}}%
{=}i(\left\vert \dot{m}\right\rangle \left\langle m\right\vert -\left\vert
m\right\rangle \left\langle \dot{m}\right\vert )$ where $\left\vert
m\right\rangle $ is given in Eq. (\ref{s5a}) with respect to the orthogonal
basis $\left\{  \left\vert m\right\rangle \text{, }\left\vert \dot
{m}\right\rangle \right\}  $ equal to $\mathrm{H}(t)\overset{\text{def}}%
{=}h_{0}\left(  t\right)  \mathbf{1}+\mathbf{h}\left(  t\right)
\cdot\mathbf{\boldsymbol{\sigma}}$. Note that, given $\phi\left(  t\right)  $
in Eq. (\ref{s5}), $\left\vert m\right\rangle $ and $\left\vert \dot
{m}\right\rangle $ are orthogonal by construction. Furthermore, we recast the
density matrix for the pure state as $\rho\left(  t\right)  \overset
{\text{def}}{=}\left\vert m(t)\right\rangle \left\langle m(t)\right\vert
=(1/2)\left[  \mathbf{1}+\mathbf{a}(t)\cdot\mathbf{\boldsymbol{\sigma}%
}\right]  $. After some algebra, we get that the Bloch vector $\mathbf{a}(t)$
and the magnetic field vector $\mathbf{h}\left(  t\right)  $ can be expressed
as%
\begin{equation}
\mathbf{a}(t)=\left(
\begin{array}
[c]{c}%
\cos\left(  \nu_{0}t\right)  \sin\left(  2\omega_{0}t\right) \\
\sin\left(  \nu_{0}t\right)  \sin\left(  2\omega_{0}t\right) \\
\cos\left(  2\omega_{0}t\right)
\end{array}
\right)  \text{,} \label{BV}%
\end{equation}
and,%
\begin{equation}
\mathbf{h}\left(  t\right)  =\left(
\begin{array}
[c]{c}%
-\frac{\nu_{0}}{2}\cos\left(  2\omega_{0}t\right)  \sin(2\omega_{0}%
t)\cos\left(  \nu_{0}t\right)  -\omega_{0}\sin\left(  \nu_{0}t\right) \\
-\frac{\nu_{0}}{2}\cos\left(  2\omega_{0}t\right)  \sin(2\omega_{0}%
t)\sin\left(  \nu_{0}t\right)  +\omega_{0}\cos\left(  \nu_{0}t\right) \\
\frac{\nu_{0}}{2}\sin^{2}\left(  2\omega_{0}t\right)
\end{array}
\right)  \text{,} \label{m2}%
\end{equation}
respectively, where $h_{0}\left(  t\right)  =0$. From Eqs. (\ref{BV}) and
(\ref{m2}), we observe that $\mathbf{a}\cdot\mathbf{h}=0$ and $\mathbf{h}%
\times\mathbf{\dot{h}\neq0}$. Therefore, the acceleration limit is not
saturated and there exists a strict inequality $a_{\mathrm{H}}<\left(
a_{\mathrm{H}}\right)  _{\max}$. More explicitly, we have%
\begin{equation}
a_{\mathrm{H}}^{2}(t)\overset{\text{def}}{=}\frac{\left(  \mathbf{h\cdot
\dot{h}}\right)  ^{2}}{\mathbf{h}^{2}}=\frac{\nu_{0}^{4}\sin^{2}(4\omega
_{0}t)}{16+4\left(  \frac{\nu_{0}}{\omega_{0}}\right)  ^{2}\sin^{2}\left(
2\omega_{0}t\right)  }\text{,} \label{care1}%
\end{equation}
and,%
\begin{equation}
\left(  a_{\mathrm{H}}(t)\right)  _{\max}^{2}=\mathcal{A}_{\max}^{2}\left(
t\right)  \overset{\text{def}}{=}\mathbf{\dot{h}}^{2}\mathbf{=}\frac{\nu
_{0}^{4}}{16}\sin^{2}(4\omega_{0}t)+2\nu_{0}^{2}\omega_{0}^{2}\left[
1+\cos(4\omega_{0}t)\right]  \text{.} \label{care2}%
\end{equation}
For more details on the quantum evolution generated by the traceless Hermitian
Hamiltonian operator $\mathrm{H}(t)\overset{\text{def}}{=}\mathbf{h}\left(
t\right)  \cdot\mathbf{\boldsymbol{\sigma}}$ specified by the magnetic field
vector $\mathbf{h}\left(  t\right)  $ in Eq. (\ref{m2}), we refer to Refs.
\cite{cafaro24A,leo24}.

\subsection{Saturation without maximum, $a_{\mathrm{H}}=\left(  a_{\mathrm{H}%
}\right)  _{\max}<\mathcal{A}_{\max}$}

In this second example, we consider a quantum evolution specified by a
time-dependent Hamiltonian given by \textrm{H}$\left(  t\right)
\overset{\text{def}}{=}\mathbf{h}\left(  t\right)  \mathbf{\cdot
\boldsymbol{\sigma}}$, with $\mathbf{h}\left(  t\right)  \overset{\text{def}%
}{=}\gamma\left(  t\right)  \hat{z}$ and $\gamma\left(  t\right)  >0$ for any
$t$. Moreover, assuming that the initial state of the system is given by
$\left\vert \psi\left(  0\right)  \right\rangle \overset{\text{def}}{=}%
(\sqrt{3}/2)|0\rangle+(1/2)|1\rangle$, the state of the system at arbitrary
time $t$ becomes%
\begin{equation}
\left\vert \psi\left(  t\right)  \right\rangle =e^{-i\left(  \int_{0}%
^{t}\gamma\left(  t^{\prime}\right)  dt^{\prime}\right)  \sigma_{z}}\left\vert
\psi\left(  0\right)  \right\rangle =\frac{\sqrt{3}}{2}e^{-i\int_{0}^{t}%
\gamma\left(  t^{\prime}\right)  dt^{\prime}}\left\vert 0\right\rangle
+\frac{1}{2}e^{i\int_{0}^{t}\gamma\left(  t^{\prime}\right)  dt^{\prime}%
}\left\vert 1\right\rangle \text{,} \label{miss}%
\end{equation}
where we have set $\hslash=1$. For the ease of presentation, we have chosen
here a specific $\left\vert \psi\left(  0\right)  \right\rangle $
parameterized by a particular choice of polar and azimuthal angles on the
Bloch sphere (i.e\textbf{.,} $\theta=\pi/3$ and $\varphi=0$, respectively).
The state $\left\vert \psi\left(  t\right)  \right\rangle $ in Eq.
(\ref{miss}) is such that $i\partial_{t}\left\vert \psi\left(  t\right)
\right\rangle =$\textrm{H}$\left(  t\right)  \left\vert \psi\left(  t\right)
\right\rangle $. From Eq. (\ref{miss}), we recast the density matrix
corresponding to the pure state $\left\vert \psi\left(  t\right)
\right\rangle $ as $\rho\left(  t\right)  \overset{\text{def}}{=}\left\vert
\psi\left(  t\right)  \right\rangle \left\langle \psi\left(  t\right)
\right\vert =(1/2)\left[  \mathbf{1}+\mathbf{a}(t)\cdot
\mathbf{\boldsymbol{\sigma}}\right]  $. After some algebra, the Bloch vector
$\mathbf{a}(t)$ reduces to%
\begin{equation}
\mathbf{a}(t)=\left(  \frac{\sqrt{3}}{2}\cos\left(  2\int_{0}^{t}\gamma\left(
t^{\prime}\right)  dt^{\prime}\right)  \text{, }\frac{\sqrt{3}}{2}\sin\left(
2\int_{0}^{t}\gamma\left(  t^{\prime}\right)  dt^{\prime}\right)  \text{,
}\frac{1}{2}\right)  \text{.} \label{miss2}%
\end{equation}
Given $\mathbf{a}(t)$ in Eq. (\ref{miss2}) and $\mathbf{h}\left(  t\right)
\overset{\text{def}}{=}\gamma\left(  t\right)  \hat{z}$, we observe that
$\mathbf{a}\cdot\mathbf{h}\neq0$ and $\mathbf{h}\times\mathbf{\dot{h}=0}$.
Therefore, the acceleration limit is saturated because $\mathbf{h}%
\times\mathbf{\dot{h}=0}$. However, the saturation value $\left(
a_{\mathrm{H}}\right)  _{\max}$ does not reach its maximum value
$\mathcal{A}_{\max}$ since $\mathbf{a}\cdot\mathbf{h}\neq0$. More explicitly,
we have%
\begin{equation}
a_{\mathrm{H}}^{2}(t)\overset{\text{def}}{=}\frac{\left[  \mathbf{h\cdot
\dot{h}-}\left(  \mathbf{a\cdot h}\right)  \left(  \mathbf{a\cdot\dot{h}%
}\right)  \right]  ^{2}}{\mathbf{h}^{2}-\left(  \mathbf{a\cdot h}\right)
^{2}}=\frac{3}{4}\dot{\gamma}^{2}\text{,} \label{care3}%
\end{equation}
and,%
\begin{equation}
\left(  a_{\mathrm{H}}(t)\right)  _{\max}^{2}\overset{\text{def}}%
{=}\mathbf{\dot{h}}^{2}-\left(  \mathbf{a\cdot\dot{h}}\right)  ^{2}%
\mathbf{=}\frac{3}{4}\dot{\gamma}^{2}<\mathcal{A}_{\max}^{2}\left(  t\right)
\overset{\text{def}}{=}\mathbf{\dot{h}}^{2}=\dot{\gamma}^{2} \label{care4}%
\end{equation}
We are now ready to present our final illustrative example.

\subsection{Saturation with maximum, $a_{\mathrm{H}}=\left(  a_{\mathrm{H}%
}\right)  _{\max}=\mathcal{A}_{\max}$}

In this third example, we consider the evolution from the state $\left\vert
\psi\left(  0\right)  \right\rangle =\left[  \left\vert 0\right\rangle
+\left\vert 1\right\rangle \right]  /\sqrt{2}$ to the state $\left\vert
\psi\left(  t\right)  \right\rangle $ under the time-dependent Hamiltonian
\textrm{H}$\left(  t\right)  \overset{\text{def}}{=}\mathbf{h}\left(
t\right)  \cdot\mathbf{\boldsymbol{\sigma}}$, with
\begin{equation}
\mathbf{h}\left(  t\right)  \overset{\text{def}}{=}\hslash\omega_{0}\cos
(\nu_{0}t)\hat{z}=\left(  0\text{, }0\text{, }\hslash\omega_{0}\cos(\nu
_{0}t)\right)  \text{.} \label{MF1}%
\end{equation}
In what follows, we set $\hslash=1$ and assume working conditions for which
the parameters $\omega_{0}$ and $\nu_{0}$ are strictly positive real
quantities. The state $\left\vert \psi\left(  t\right)  \right\rangle
=e^{-i\int_{0}^{t}\mathrm{H}\left(  t^{\prime}\right)  dt^{\prime}}\left\vert
\psi\left(  0\right)  \right\rangle $ is given by,%
\begin{equation}
\left\vert \psi\left(  t\right)  \right\rangle =e^{-i\left[  \frac{\omega_{0}%
}{\nu_{0}}\sin(\nu_{0}t)\right]  }\left(
\begin{array}
[c]{c}%
\frac{1}{\sqrt{2}}\\
\frac{e^{2i\left[  \frac{\omega_{0}}{\nu_{0}}\sin(\nu_{0}t)\right]  }}%
{\sqrt{2}}%
\end{array}
\right)  \simeq\left(
\begin{array}
[c]{c}%
\frac{1}{\sqrt{2}}\\
\frac{e^{2i\left[  \frac{\omega_{0}}{\nu_{0}}\sin(\nu_{0}t)\right]  }}%
{\sqrt{2}}%
\end{array}
\right)  \text{,} \label{mil1}%
\end{equation}
where the symbol \textquotedblleft$\simeq$\textquotedblright\ in Eq.
(\ref{mil1}) denotes physical equivalence of pure quantum states under global
phase factors. From Eq. (\ref{mil1}), we express the density matrix that
corresponds to the pure state $\left\vert \psi\left(  t\right)  \right\rangle
$ in Eq. (\ref{mil1}) as $\rho\left(  t\right)  \overset{\text{def}}%
{=}\left\vert \psi\left(  t\right)  \right\rangle \left\langle \psi\left(
t\right)  \right\vert =(1/2)\left[  \mathbf{1}+\mathbf{a}(t)\cdot
\mathbf{\boldsymbol{\sigma}}\right]  $. After some algebraic manipulation, the
Bloch vector $\mathbf{a}(t)$ becomes%
\begin{equation}
\mathbf{a}\left(  t\right)  =\left(  \cos\left[  2\frac{\omega_{0}}{\nu_{0}%
}\sin(\nu_{0}t)\right]  \text{, }\sin\left[  2\frac{\omega_{0}}{\nu_{0}}%
\sin(\nu_{0}t)\right]  \text{, }0\right)  \text{.} \label{BV1}%
\end{equation}
From Eqs. (\ref{MF1}) and (\ref{BV1}), we observe that $\mathbf{a}%
\cdot\mathbf{h}=0$ and $\mathbf{h}\times\mathbf{\dot{h}}=\mathbf{0}$.
Therefore, the acceleration limit is saturated because $\mathbf{h}%
\times\mathbf{\dot{h}=0}$. Moreover, the saturation value $\left(
a_{\mathrm{H}}\right)  _{\max}$ assumes its maximum value $\mathcal{A}_{\max}$
since $\mathbf{a}\cdot\mathbf{h}=0$. More explicitly, we have%
\begin{equation}
a_{\mathrm{H}}^{2}(t)\overset{\text{def}}{=}\frac{\left(  \mathbf{h\cdot
\dot{h}}\right)  ^{2}}{\mathbf{h}^{2}}=\omega_{0}^{2}\nu_{0}^{2}\sin^{2}%
(\nu_{0}t)\text{,} \label{care5}%
\end{equation}
and,%
\begin{equation}
\left(  a_{\mathrm{H}}(t)\right)  _{\max}^{2}=\mathcal{A}_{\max}^{2}\left(
t\right)  \overset{\text{def}}{=}\mathbf{\dot{h}}^{2}=\omega_{0}^{2}\nu
_{0}^{2}\sin^{2}(\nu_{0}t)\text{.} \label{care6}%
\end{equation}
For a schematic depiction of the geometric conditions that specify the three
scenarios considered in our illustrative examples, we refer to Fig.
$1$.\begin{figure}[t]
\centering
\includegraphics[width=0.75\textwidth] {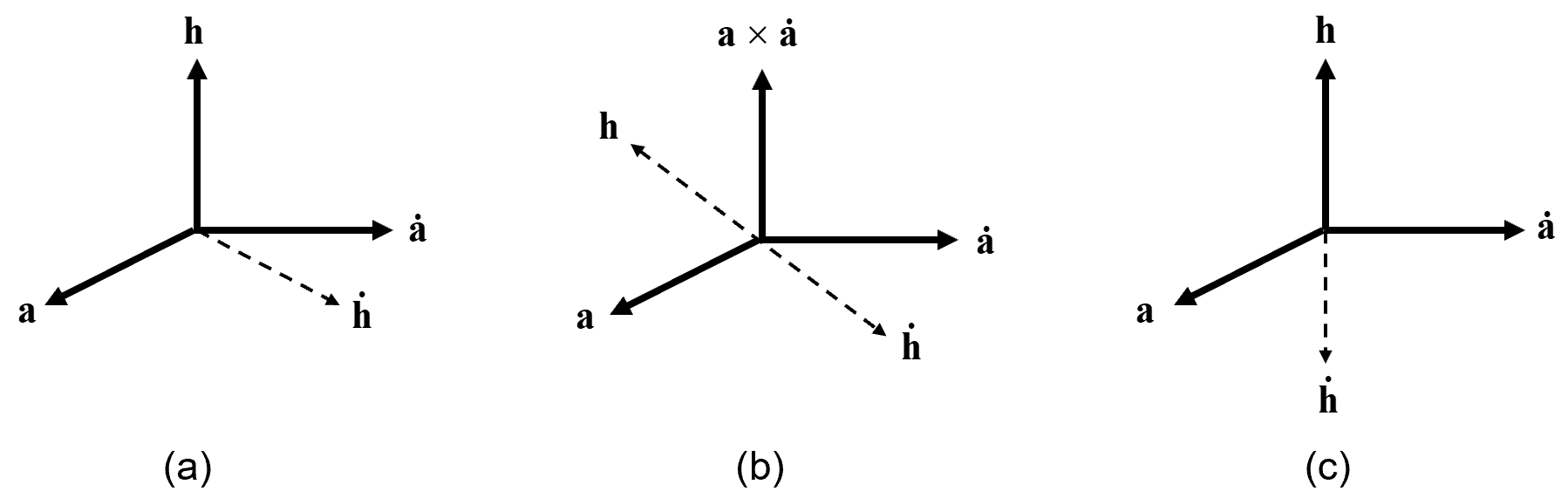}\caption{Visual sketch of the
geometric conditions that specify the three scenarios considered in our
illustrative examples. In (a), since $\mathbf{h}$ and $\mathbf{\dot{h}}$ are
not collinear, $a_{\mathrm{H}}<\left(  a_{\mathrm{H}}\right)  _{\max}$ (i.e.,
no saturation). In (b), $\mathbf{h}$ and $\mathbf{\dot{h}}$ are collinear.
However, $\mathbf{a}$ and $\mathbf{h}$ are not orthogonal. Therefore, we have
in this case $a_{\mathrm{H}}=\left(  a_{\mathrm{H}}\right)  _{\max
}<\mathcal{A}_{\max}$ (i.e., saturation without maximum). Finally, not only
$\mathbf{h}$ and $\mathbf{\dot{h}}$ are collinear in (c). In this scenario, we
also have that $\mathbf{a}$ and $\mathbf{h}$ are orthogonal. Therefore, in
this case $a_{\mathrm{H}}=\left(  a_{\mathrm{H}}\right)  _{\max}%
=\mathcal{A}_{\max}$ (i.e., saturation with maximum). Mutually orthogonal
vectors are in bold.}%
\end{figure}Before discussing our final remarks, we note that Pati reported in
Ref. \cite{pati23} that the quantum acceleration limit saturates in any
finite-dimensional Hilbert space where the nonstationary Hamiltonian of the
system is given by \textrm{H}$\left(  t\right)  \overset{\text{def}}{=}%
\lambda\left(  t\right)  \mathrm{H}_{0}$. Here, $\mathrm{H}_{0}$ is a
time-independent operator, while $\lambda\left(  t\right)  $ denotes a
time-varying coupling parameter. In our geometric picture for single-qubit
evolutions, Pati's result can be simply explained by noting that since
\textrm{H}$\left(  t\right)  \overset{\text{def}}{=}\lambda\left(  t\right)
\hat{h}\cdot\mathbf{\sigma}$ with $\hat{h}$ constant in time, the acceleration
limit saturates since $\mathbf{h}\left(  t\right)  \overset{\text{def}}%
{=}\lambda\left(  t\right)  \hat{h}$ and $\mathbf{\dot{h}}\left(  t\right)  $
are collinear vectors.

We are now ready for our summary of results and final comments.

\section{Conclusion}

In this paper, we began with a comparative analysis between the derivations of
the quantum acceleration limit for arbitrary finite-dimensional quantum
systems in a pure state as originally proposed by Pati (Eq. (\ref{luo5})) and
Alsing-Cafaro (Eq. (\ref{cicca})). The visual summary of this analysis appears
in Table I. Then, limiting our attention to the quantum evolution of two-level
quantum systems on a Bloch sphere under general time-dependent Hamiltonians,
we recast (Eq. (\ref{king3})) the acceleration limit inequality by means of
the Bloch vector $\mathbf{a}\left(  t\right)  $ of the system and of the
magnetic field vector $\mathbf{h}\left(  t\right)  $ that characterizes the
(traceless) Hermitian Hamiltonian of the system \textrm{H}$\left(  t\right)
\overset{\text{def}}{=}\mathbf{h}\left(  t\right)  \cdot
\mathbf{\boldsymbol{\sigma}}$. We then find the proper geometric conditions in
terms of $\mathbf{a}\left(  t\right)  $, $\mathbf{h}\left(  t\right)  $, and
their temporal derivatives for which the above-mentioned inequality is
saturated and the quantum acceleration is maximum. A visual summary of these
conditions appears in Table II. To illustrate these conditions in an explicit
manner, we then discuss three illustrative examples. In the first scenario,
the quantum acceleration $a_{\mathrm{H}}\left(  t\right)  $ (Eq.
(\ref{care1})) does not reach the value of its upper bound $\left(
a_{\mathrm{H}}\left(  t\right)  \right)  _{\max}$ (Eq. (\ref{care2})) (i.e.,
no saturation since $a_{\mathrm{H}}\left(  t\right)  <\left(  a_{\mathrm{H}%
}\left(  t\right)  \right)  _{\max}$). This first scenario is specified by the
relations $\mathbf{a}\left(  t\right)  \cdot\mathbf{h}\left(  t\right)  =0$
and $\mathbf{h}\left(  t\right)  \times\mathbf{\dot{h}}\left(  t\right)
\neq\mathbf{0}$. In the second scenario, the quantum acceleration
$a_{\mathrm{H}}\left(  t\right)  $ (Eq. (\ref{care3})) reaches the value of
its upper bound $\left(  a_{\mathrm{H}}\left(  t\right)  \right)  _{\max}$
(Eq. (\ref{care4})). However, the upper bound $\left(  a_{\mathrm{H}}\left(
t\right)  \right)  _{\max}$ does not assumes its maximum value $\mathcal{A}%
_{\max}\left(  t\right)  $ (Eq. (\ref{care4})) (i.e., saturation without
maximum, $a_{\mathrm{H}}\left(  t\right)  =\left(  a_{\mathrm{H}}\left(
t\right)  \right)  _{\max}<\mathcal{A}_{\max}\left(  t\right)  $). This second
scenario is characterized by the conditions $\mathbf{a}\left(  t\right)
\cdot\mathbf{h}\left(  t\right)  \neq0$ and $\mathbf{h}\left(  t\right)
\times\mathbf{\dot{h}}\left(  t\right)  =\mathbf{0}$. In the third scenario,
the quantum acceleration $a_{\mathrm{H}}\left(  t\right)  $ (Eq.
(\ref{care5})) reaches the value of its upper bound $\left(  a_{\mathrm{H}%
}\left(  t\right)  \right)  _{\max}$ (Eq. (\ref{care6})). Moreover, the upper
bound $\left(  a_{\mathrm{H}}\left(  t\right)  \right)  _{\max}$ achieves its
maximum value $\mathcal{A}_{\max}\left(  t\right)  $ (Eq. (\ref{care6}))
(i.e., saturation with maximum, $a_{\mathrm{H}}\left(  t\right)  =\left(
a_{\mathrm{H}}\left(  t\right)  \right)  _{\max}=\mathcal{A}_{\max}\left(
t\right)  $). Finally, this third scenario is described by the relations
$\mathbf{a}\left(  t\right)  \cdot\mathbf{h}\left(  t\right)  =0$ and
$\mathbf{h}\left(  t\right)  \times\mathbf{\dot{h}}\left(  t\right)
=\mathbf{0}$.

\medskip

In this paper, the presentation of our illustrative examples was limited to
two-level quantum systems in a pure state, despite the fact that our formal
derivation of the quantum acceleration limit applies to any finite-dimensional
quantum system in a pure state. Therefore, two clear extensions of our work is
its explicit application to higher-dimensional systems in a pure state and, in
addition, its conceptual generalization to quantum systems in a mixed quantum
state. In this basic single-qubit scenario, unlike what occurs in
higher-dimensional cases, the notions of Bloch vectors and Bloch spheres have
both a neat physical significance and a clear geometric visualization. When
shifting from two-level quantum systems to higher dimensional systems, the
physical meaning of the Bloch vector conserves its usefulness. However, its
geometric interpretation is less transparent than that achieved for
single-qubit quantum systems
\cite{jakob01,kimura03,krammer08,kurzy11,xie20,siewert21}. We refer to Ref.
\cite{gamel16} for a very helpful discussion on the Bloch vector
representations of single-qubit systems, single-qutrit systems, and two-qubit
systems by means of Pauli, Gell-Mann, and Dirac matrices, respectively.

\medskip

Moreover, when transitioning from pure states to mixed quantum states, we have
to tackle additional challenges even for lower-dimensional quantum systems.
One of the main obstacles (absent for pure states) is the existence of
infinitely many distinguishability distances for mixed quantum states
\cite{bures69,uhlman76,hubner92,karol06,erik20,alsingpra23,chi24}. This
freedom in the selection of the metric leads to the conclusion that geometric
investigations of physical phenomena are still open to metric-dependent
interpretations. Despite much progress, given the non-uniqueness of
distinguishability distances for mixed states, comprehending the physical
importance of employing either metric remains an objective of significant
theoretical and applied relevance \cite{silva21,silva21B,mera22}.

\medskip

Although our work is purely driven by conceptual interests in quantum physics,
we are aware of the fact that the concept of quantum acceleration limits can
lead to practical applications in quantum technology. As a matter of fact,
upper limits on the speed, the acceleration, and the jerk (i.e., the rate of
change of the acceleration) of a quantum evolution can be employed to quantify
the complexity and the efficiency of quantum processes
\cite{uzdin12,campaioli19,cafaroQR,carloprd22,carlopre22,carlocqg23}, to gauge
the cost of controlled quantum evolutions \cite{frey16,deffner17,saito23}, and
to characterize losses when quantum critical behaviors are present
\cite{liu19,matt23,araki23}. For the time being, we leave the study of some of
these applied aspects of our research to future scientific endeavours.

\begin{acknowledgments}
C.C. acknowledges interesting discussions on quantum acceleration limits with
Paul M. Alsing, Selman Ipek, and Biswaranjan Panda. C.C. is also thankful to
Ryusuke Hamazaki and Chryssomalis Chryssomalakos for bringing to his attention
closely related findings reported in Refs. \cite{hamazaki23,hamazaki22} and
Refs. \cite{flores23,flores24}, respectively. Furthermore, the authors are
grateful to anonymous referees for constructive comments leading to an
improved version of the paper. Any opinions, findings and conclusions or
recommendations expressed in this material are those of the author(s) and do
not necessarily reflect the views of their home Institutions.
\end{acknowledgments}

\pagebreak

\appendix

\section{Derivation of the inequality in Eq. (\ref{SR})}

In this Appendix, we present a derivation of the so-called
Robertson-Schr\"{o}dinger uncertainty relation in Eq. (\ref{SR})
\cite{robertson,schrodinger}. For an interesting approach to proving
Robertson-type uncertainty relations using the concept of quantum Fisher
information, we refer to Ref. \cite{gibilisco08}. For a discussion on stronger
uncertainty relations for all incompatible observables in quantum mechanics,
relating to the sum of variances, we suggest Ref. \cite{maccone14}.

Let us consider two Hermitian operators $A$ and $B$. Then, the
Robertson-Schr\"{o}dinger uncertainty relation states that%
\begin{equation}
\sigma_{A}^{2}\sigma_{B}^{2}\geq\left\vert \frac{1}{2}\left\langle \left\{
A\text{, }B\right\}  \right\rangle -\left\langle A\right\rangle \left\langle
B\right\rangle \right\vert ^{2}+\left\vert \frac{1}{2i}\left\langle \left[
A\text{, }B\right]  \right\rangle \right\vert ^{2}\text{,} \label{ines1}%
\end{equation}
where $\sigma_{A}^{2}\overset{\text{def}}{=}\left\langle \left(
A-\left\langle A\right\rangle \right)  ^{2}\right\rangle =\left\langle
A^{2}\right\rangle -\left\langle A\right\rangle ^{2}$, $\left\langle
A\right\rangle \overset{\text{def}}{=}\left\langle \psi\left\vert A\right\vert
\psi\right\rangle $, $\left\{  A\text{, }B\right\}  \overset{\text{def}}%
{=}AB+BA$, and $\left[  A\text{, }B\right]  \overset{\text{def}}{=}AB-BA$. A
proof of the inequality in Eq. (\ref{ines1}) is as follows. Note that,%
\begin{align}
\sigma_{A}^{2}  &  =\left\langle \psi\left\vert \left(  A-\left\langle
A\right\rangle \right)  ^{2}\right\vert \psi\right\rangle \nonumber\\
&  =\left\langle \psi\left\vert \left(  A-\left\langle A\right\rangle \right)
\left(  A-\left\langle A\right\rangle \right)  \right\vert \psi\right\rangle
\nonumber\\
&  =\left\langle f\left\vert f\right.  \right\rangle \text{,}%
\end{align}
with $\left\vert f\right\rangle \overset{\text{def}}{=}\left(  A-\left\langle
A\right\rangle \right)  \left\vert \psi\right\rangle $. Similarly, $\sigma
_{B}^{2}=\left\langle g\left\vert g\right.  \right\rangle $ with $\left\vert
g\right\rangle \overset{\text{def}}{=}\left(  B-\left\langle B\right\rangle
\right)  \left\vert \psi\right\rangle $. Therefore, $\sigma_{A}^{2}\sigma
_{B}^{2}=\left\langle f\left\vert f\right.  \right\rangle \left\langle
g\left\vert g\right.  \right\rangle $. Using the Cauchy-Schwarz inequality,
$\left\vert \left\langle f\left\vert g\right.  \right\rangle \right\vert
^{2}\leq\left\langle f\left\vert f\right.  \right\rangle \left\langle
g\left\vert g\right.  \right\rangle $, we have that%
\begin{equation}
\sigma_{A}^{2}\sigma_{B}^{2}\geq\left\vert \left\langle f\left\vert g\right.
\right\rangle \right\vert ^{2}\text{.} \label{04}%
\end{equation}
Observe that $\left\langle f\left\vert g\right.  \right\rangle \in%
\mathbb{C}
$, in general. Then, we have
\begin{align}
\left\vert \left\langle f\left\vert g\right.  \right\rangle \right\vert ^{2}
&  =\left\langle f\left\vert g\right.  \right\rangle \left\langle f\left\vert
g\right.  \right\rangle ^{\ast}\nonumber\\
&  =\left\langle f\left\vert g\right.  \right\rangle \left\langle g\left\vert
f\right.  \right\rangle \nonumber\\
&  =\left[  \operatorname{Re}\left(  \left\langle f\left\vert g\right.
\right\rangle \right)  \right]  ^{2}+\left[  \operatorname{Im}\left(
\left\langle f\left\vert g\right.  \right\rangle \right)  \right]
^{2}\nonumber\\
&  =\left(  \frac{\left\langle f\left\vert g\right.  \right\rangle
+\left\langle g\left\vert f\right.  \right\rangle }{2}\right)  ^{2}+\left(
\frac{\left\langle f\left\vert g\right.  \right\rangle -\left\langle
g\left\vert f\right.  \right\rangle }{2i}\right)  ^{2}\text{,}%
\end{align}
that is,%
\begin{equation}
\left\vert \left\langle f\left\vert g\right.  \right\rangle \right\vert
^{2}=\left(  \frac{\left\langle f\left\vert g\right.  \right\rangle
+\left\langle g\left\vert f\right.  \right\rangle }{2}\right)  ^{2}+\left(
\frac{\left\langle f\left\vert g\right.  \right\rangle -\left\langle
g\left\vert f\right.  \right\rangle }{2i}\right)  ^{2}\text{.} \label{a10}%
\end{equation}
Since $A$ and $B$ are Hermitian operators, we have
\begin{align}
\left\langle f\left\vert g\right.  \right\rangle  &  =\left\langle
\psi\left\vert \left(  A-\left\langle A\right\rangle \right)  \left(
B-\left\langle B\right\rangle \right)  \right\vert \psi\right\rangle
\nonumber\\
&  =\left\langle \psi\left\vert AB-\left\langle B\right\rangle A-\left\langle
A\right\rangle B-\left\langle A\right\rangle \left\langle B\right\rangle
\right\vert \psi\right\rangle \nonumber\\
&  =\left\langle AB\right\rangle -\left\langle A\right\rangle \left\langle
B\right\rangle \text{,}%
\end{align}
that is,%
\begin{equation}
\left\langle f\left\vert g\right.  \right\rangle =\left\langle AB\right\rangle
-\left\langle A\right\rangle \left\langle B\right\rangle \text{.} \label{a11}%
\end{equation}
Similarly, we have%
\begin{equation}
\left\langle g\left\vert f\right.  \right\rangle =\left\langle BA\right\rangle
-\left\langle A\right\rangle \left\langle B\right\rangle \text{.} \label{a12}%
\end{equation}
Combining Eqs. (\ref{a11}) and (\ref{a12}), we get%
\begin{align}
\left\langle f\left\vert g\right.  \right\rangle +\left\langle g\left\vert
f\right.  \right\rangle  &  =\left\langle AB\right\rangle -\left\langle
A\right\rangle \left\langle B\right\rangle +\left\langle BA\right\rangle
-\left\langle A\right\rangle \left\langle B\right\rangle \nonumber\\
&  =\left\langle AB+BA\right\rangle -2\left\langle A\right\rangle \left\langle
B\right\rangle \nonumber\\
&  =\left\langle \left\{  A\text{, }B\right\}  \right\rangle -2\left\langle
A\right\rangle \left\langle B\right\rangle \text{,}%
\end{align}
that is,%
\begin{equation}
\left\langle f\left\vert g\right.  \right\rangle +\left\langle g\left\vert
f\right.  \right\rangle =\left\langle \left\{  A\text{, }B\right\}
\right\rangle -2\left\langle A\right\rangle \left\langle B\right\rangle
\text{.} \label{a13}%
\end{equation}
Furthermore, we obtain%
\begin{align}
\left\langle f\left\vert g\right.  \right\rangle -\left\langle g\left\vert
f\right.  \right\rangle  &  =\left\langle AB\right\rangle -\left\langle
A\right\rangle \left\langle B\right\rangle -\left\langle BA\right\rangle
+\left\langle A\right\rangle \left\langle B\right\rangle \nonumber\\
&  =\left\langle AB-BA\right\rangle \nonumber\\
&  =\left\langle \left[  A\text{, }B\right]  \right\rangle \text{,}%
\end{align}
that is,%
\begin{equation}
\left\langle f\left\vert g\right.  \right\rangle -\left\langle g\left\vert
f\right.  \right\rangle =\left\langle \left[  A\text{, }B\right]
\right\rangle \text{.} \label{a14}%
\end{equation}
Employing Eqs. (\ref{a13}) and (\ref{a14}), Eq. (\ref{a10}) reduces to%
\begin{equation}
\left\vert \left\langle f\left\vert g\right.  \right\rangle \right\vert
^{2}=\left(  \frac{1}{2}\left\langle \left\{  A\text{, }B\right\}
\right\rangle -\left\langle A\right\rangle \left\langle B\right\rangle
\right)  ^{2}+\left(  \frac{1}{2i}\left\langle \left[  A\text{, }B\right]
\right\rangle \right)  ^{2}\text{.} \label{a15}%
\end{equation}
Finally, using Eq. (\ref{a15}), Eq. (\ref{04}) becomes%
\begin{equation}
\sigma_{A}^{2}\sigma_{B}^{2}\geq\left\vert \frac{1}{2}\left\langle \left\{
A\text{, }B\right\}  \right\rangle -\left\langle A\right\rangle \left\langle
B\right\rangle \right\vert ^{2}+\left\vert \frac{1}{2i}\left\langle \left[
A\text{, }B\right]  \right\rangle \right\vert ^{2}\text{,} \label{a16}%
\end{equation}
that is,%
\begin{equation}
\sigma_{A}^{2}\sigma_{B}^{2}\geq\frac{1}{4}\left\vert \left\langle \left[
A\text{, }B\right]  \right\rangle \right\vert ^{2}+\left\vert \mathrm{cov}%
\left(  A\text{, }B\right)  \right\vert ^{2}\geq\frac{1}{4}\left\vert
\left\langle \left[  A\text{, }B\right]  \right\rangle \right\vert
^{2}\text{,} \label{clock1}%
\end{equation}
where $\mathrm{cov}\left(  A\text{, }B\right)  \overset{\text{def}}%
{=}\left\langle \frac{AB+BA}{2}\right\rangle -\left\langle A\right\rangle
\left\langle B\right\rangle $ is the covariance of the two observables $A$ and
$B$. For completeness, observe that $\left[  A\text{, }B\right]  $ is a
anti-Hermitian operator. Thus, $\left\langle \left[  A\text{, }B\right]
\right\rangle $ is a purely imaginary number. The derivation of the first
inequality in Eq. (\ref{clock1}) ends our demonstration. Moreover, the second
(less tight) inequality in Eq. (\ref{clock1}), $\sigma_{A}^{2}\sigma_{B}%
^{2}\geq(1/4)\left\vert \left\langle \left[  A\text{, }B\right]  \right\rangle
\right\vert ^{2}$, denotes the Robertson uncertainty relation. Finally, for
more rigorous mathematical details on the derivation of the inequality in Eq.
(\ref{a16}), we suggest Ref. \cite{hall13}.

\section{Derivations of Eqs. (\ref{sigma1}) and (\ref{sigma3})}

In this Appendix, using standard techniques in quantum computing
\cite{nielsen00,polak11,hidary19}, we offer a derivation of Eqs.
(\ref{sigma1}) and (\ref{sigma3}).

\subsection{Deriving Eq. (\ref{sigma1})}

We want to derive here the relation $\sigma_{\mathrm{H}}^{2}\overset
{\text{def}}{=}\mathrm{tr}\left(  \rho\mathrm{H}^{2}\right)  -\left[
\mathrm{tr}\left(  \rho\mathrm{H}\right)  \right]  ^{2}=\mathbf{h}^{2}-\left(
\mathbf{a}\cdot\mathbf{h}\right)  ^{2}$ exploiting the fact that $\rho\left(
t\right)  \overset{\text{def}}{=}(1/2)\left(  \mathbf{1+a\cdot
\boldsymbol{\sigma}}\right)  $ and \textrm{H}$\left(  t\right)  \overset
{\text{def}}{=}$ $h_{0}\mathbf{1+h}\cdot\mathbf{\boldsymbol{\sigma}}$. For
generality, we are assuming $h_{0}=h_{0}\left(  t\right)  \neq0$. We begin by
observing that,%
\begin{align}
\mathrm{tr}\left(  \rho\mathrm{H}\right)   &  =\mathrm{tr}\left[  \left(
\frac{\mathbf{1}+\mathbf{a}\cdot\mathbf{\boldsymbol{\sigma}}}{2}\right)
\left(  h_{0}\mathbf{1}+\mathbf{h}\cdot\mathbf{\boldsymbol{\sigma}}\right)
\right] \nonumber\\
&  =\mathrm{tr}\left[  \left(  \frac{\mathbf{1}}{2}+\frac{\mathbf{a}%
\cdot\mathbf{\boldsymbol{\sigma}}}{2}\right)  \left(  h_{0}\mathbf{1}%
+\mathbf{h}\cdot\mathbf{\boldsymbol{\sigma}}\right)  \right] \nonumber\\
&  =\mathrm{tr}\left[  \frac{h_{0}}{2}\mathbf{1}+\frac{1}{2}\mathbf{h}%
\cdot\mathbf{\boldsymbol{\sigma}}+\frac{h_{0}}{2}\mathbf{a}\cdot
\mathbf{\boldsymbol{\sigma}}+\frac{1}{2}\left(  \mathbf{a}\cdot
\mathbf{\boldsymbol{\sigma}}\right)  \left(  \mathbf{h}\cdot
\mathbf{\boldsymbol{\sigma}}\right)  \right] \nonumber\\
&  =\frac{h_{0}}{2}\mathrm{tr}\left(  \mathbf{1}\right)  +\frac{1}%
{2}\mathrm{tr}\left(  \mathbf{h}\cdot\mathbf{\boldsymbol{\sigma}}\right)
+\frac{h_{0}}{2}\mathrm{tr}\left(  \mathbf{a}\cdot\mathbf{\boldsymbol{\sigma}%
}\right)  +\frac{1}{2}\mathrm{tr}\left[  \left(  \mathbf{a}\cdot
\mathbf{\boldsymbol{\sigma}}\right)  \left(  \mathbf{h}\cdot
\mathbf{\boldsymbol{\sigma}}\right)  \right]  \text{,}%
\end{align}
that is,%
\begin{equation}
\mathrm{tr}\left(  \rho\mathrm{H}\right)  =h_{0}+\mathbf{a}\cdot
\mathbf{h}\text{,} \label{r3}%
\end{equation}
since $\mathrm{tr}\left(  \mathbf{1}\right)  =2$, $\mathrm{tr}\left(
\mathbf{h}\cdot\mathbf{\boldsymbol{\sigma}}\right)  =0$, $\mathrm{tr}\left(
\mathbf{a}\cdot\mathbf{\boldsymbol{\sigma}}\right)  =0$, and $\mathrm{tr}%
\left[  \left(  \mathbf{a}\cdot\mathbf{\boldsymbol{\sigma}}\right)  \left(
\mathbf{h}\cdot\mathbf{\boldsymbol{\sigma}}\right)  \right]  =2\mathbf{a}%
\cdot\mathbf{h}$. Let us focus now on the calculation of $\mathrm{tr}\left(
\rho\mathrm{H}^{2}\right)  $. We have,%
\begin{align}
\rho\mathrm{H}^{2}  &  =\left(  \frac{\mathbf{1}+\mathbf{a}\cdot
\mathbf{\boldsymbol{\sigma}}}{2}\right)  \left(  h_{0}\mathbf{1}%
+\mathbf{h}\cdot\mathbf{\boldsymbol{\sigma}}\right)  ^{2}\nonumber\\
&  =\left(  \frac{\mathbf{1}}{2}+\frac{\mathbf{a}\cdot
\mathbf{\boldsymbol{\sigma}}}{2}\right)  \left[  h_{0}^{2}\mathbf{1}+\left(
\mathbf{h}\cdot\mathbf{\boldsymbol{\sigma}}\right)  ^{2}+2h_{0}\left(
\mathbf{h}\cdot\mathbf{\boldsymbol{\sigma}}\right)  \right] \nonumber\\
&  =\left[  \frac{h_{0}^{2}}{2}\mathbf{1}+\frac{1}{2}\left(  \mathbf{h}%
\cdot\mathbf{\boldsymbol{\sigma}}\right)  ^{2}+h_{0}\left(  \mathbf{h}%
\cdot\mathbf{\boldsymbol{\sigma}}\right)  +\frac{h_{0}^{2}}{2}\left(
\mathbf{a}\cdot\mathbf{\boldsymbol{\sigma}}\right)  +\frac{1}{2}\left(
\mathbf{a}\cdot\mathbf{\boldsymbol{\sigma}}\right)  \left(  \mathbf{h}%
\cdot\mathbf{\boldsymbol{\sigma}}\right)  ^{2}+h_{0}\left(  \mathbf{a}%
\cdot\mathbf{\boldsymbol{\sigma}}\right)  \left(  \mathbf{h}\cdot
\mathbf{\boldsymbol{\sigma}}\right)  \right]  \text{,}%
\end{align}
that is,%
\begin{equation}
\mathrm{tr}\left(  \rho\mathrm{H}^{2}\right)  =h_{0}^{2}+\left\Vert
\mathbf{h}\right\Vert ^{2}+2h_{0}\left(  \mathbf{a}\cdot\mathbf{h}\right)
\text{,} \label{r4}%
\end{equation}
since $\mathrm{tr}\left(  \mathbf{1}\right)  =2$, $\mathrm{tr}\left[  \left(
\mathbf{h}\cdot\mathbf{\boldsymbol{\sigma}}\right)  ^{2}\right]  =2\left\Vert
\mathbf{h}\right\Vert ^{2}$, $\mathrm{tr}\left(  \mathbf{h}\cdot
\mathbf{\boldsymbol{\sigma}}\right)  =0$, $\mathrm{tr}\left(  \mathbf{a}%
\cdot\mathbf{\boldsymbol{\sigma}}\right)  =0$, $\mathrm{tr}\left[  \left(
\mathbf{a}\cdot\mathbf{\boldsymbol{\sigma}}\right)  \left(  \mathbf{h}%
\cdot\mathbf{\boldsymbol{\sigma}}\right)  ^{2}\right]  =0$ and, $\mathrm{tr}%
\left[  \left(  \mathbf{a}\cdot\mathbf{\boldsymbol{\sigma}}\right)  \left(
\mathbf{h}\cdot\mathbf{\boldsymbol{\sigma}}\right)  \right]  =2\mathbf{a}%
\cdot\mathbf{h}$. Using Eqs. (\ref{r3}) and (\ref{r4}), we finally arrive at
$\sigma_{\mathrm{H}}^{2}=\mathbf{h}^{2}-\left(  \mathbf{a}\cdot\mathbf{h}%
\right)  ^{2}$.

\subsection{Deriving Eq. (\ref{sigma3})}

We wish to obtain the relation $\left\vert \left\langle \left[  \mathrm{H}%
\text{, }\mathrm{\dot{H}}\right]  \right\rangle \right\vert ^{2}=\left[
\left(  \mathbf{a\times h}\right)  \cdot\mathbf{\dot{h}-}\left(
\mathbf{a\times\dot{h}}\right)  \cdot\mathbf{h}\right]  ^{2}$ in Eq.
(\ref{sigma3}). Recalling that \textrm{H}$\overset{\text{def}}{=}$
$h_{0}\mathbf{1+h}\cdot\mathbf{\boldsymbol{\sigma}}$ and \textrm{\.{H}%
}$\overset{\text{def}}{=}$ $\dot{h}_{0}\mathbf{1+\dot{h}}\cdot
\mathbf{\boldsymbol{\sigma}}$ since we assume for generality that $h_{0}%
=h_{0}\left(  t\right)  \neq0$, we begin by noting that%
\begin{align}
\left[  \mathrm{H}\text{, }\mathrm{\dot{H}}\right]   &  =\left[
h_{0}\mathbf{1+h}\cdot\mathbf{\boldsymbol{\sigma}}\text{, }\dot{h}%
_{0}\mathbf{1+\dot{h}}\cdot\mathbf{\boldsymbol{\sigma}}\right] \nonumber\\
&  =\left[  h_{0}\mathbf{1}\text{, }\dot{h}_{0}\mathbf{1+\dot{h}}%
\cdot\mathbf{\boldsymbol{\sigma}}\right]  +\left[  \mathbf{h}\cdot
\mathbf{\boldsymbol{\sigma}}\text{, }\dot{h}_{0}\mathbf{1+\dot{h}}%
\cdot\mathbf{\boldsymbol{\sigma}}\right] \nonumber\\
&  =\left[  h_{0}\mathbf{1}\text{, }\dot{h}_{0}\mathbf{1}\right]  +\left[
h_{0}\mathbf{1}\text{, }\mathbf{\dot{h}}\cdot\mathbf{\boldsymbol{\sigma}%
}\right]  +\left[  \mathbf{h}\cdot\mathbf{\boldsymbol{\sigma}}\text{, }\dot
{h}_{0}\mathbf{1}\right]  +\left[  \mathbf{h}\cdot\mathbf{\boldsymbol{\sigma}%
}\text{, }\mathbf{\dot{h}}\cdot\mathbf{\boldsymbol{\sigma}}\right] \nonumber\\
&  =\left[  \mathbf{h}\cdot\mathbf{\boldsymbol{\sigma}}\text{, }%
\mathbf{\dot{h}}\cdot\mathbf{\boldsymbol{\sigma}}\right] \nonumber\\
&  =\left(  \mathbf{h}\cdot\mathbf{\boldsymbol{\sigma}}\right)  \left(
\mathbf{\dot{h}}\cdot\mathbf{\boldsymbol{\sigma}}\right)  -\left(
\mathbf{\dot{h}}\cdot\mathbf{\boldsymbol{\sigma}}\right)  \left(
\mathbf{h}\cdot\mathbf{\boldsymbol{\sigma}}\right)  \text{,}%
\end{align}
that is,%
\begin{equation}
\left\langle \left[  \mathrm{H}\text{, }\mathrm{\dot{H}}\right]  \right\rangle
=\left\langle \left(  \mathbf{h}\cdot\mathbf{\boldsymbol{\sigma}}\right)
\left(  \mathbf{\dot{h}}\cdot\mathbf{\boldsymbol{\sigma}}\right)  -\left(
\mathbf{\dot{h}}\cdot\mathbf{\boldsymbol{\sigma}}\right)  \left(
\mathbf{h}\cdot\mathbf{\boldsymbol{\sigma}}\right)  \right\rangle \text{.}
\label{tyler1}%
\end{equation}
From Eq. (\ref{tyler1}), we have that $\left\langle \left[  \mathrm{H}\text{,
}\mathrm{\dot{H}}\right]  \right\rangle $ becomes%
\begin{align}
\left\langle \left[  \mathrm{H}\text{, }\mathrm{\dot{H}}\right]
\right\rangle  &  =\mathrm{tr}\left(  \rho\left[  \mathrm{H}\text{,
}\mathrm{\dot{H}}\right]  \right) \nonumber\\
&  =\mathrm{tr}\left[  \left(  \frac{\mathbf{1+a}\cdot
\mathbf{\boldsymbol{\sigma}}}{2}\right)  \left[  \mathrm{H}\text{,
}\mathrm{\dot{H}}\right]  \right] \nonumber\\
&  =\mathrm{tr}\left\{  \left(  \frac{\mathbf{1+a}\cdot
\mathbf{\boldsymbol{\sigma}}}{2}\right)  \left[  \left(  \mathbf{h}%
\cdot\mathbf{\boldsymbol{\sigma}}\right)  \left(  \mathbf{\dot{h}}%
\cdot\mathbf{\boldsymbol{\sigma}}\right)  -\left(  \mathbf{\dot{h}}%
\cdot\mathbf{\boldsymbol{\sigma}}\right)  \left(  \mathbf{h}\cdot
\mathbf{\boldsymbol{\sigma}}\right)  \right]  \right\} \nonumber\\
&  =\frac{1}{2}\mathrm{tr}\left[  \left(  \mathbf{h}\cdot
\mathbf{\boldsymbol{\sigma}}\right)  \left(  \mathbf{\dot{h}}\cdot
\mathbf{\boldsymbol{\sigma}}\right)  \right]  -\frac{1}{2}\mathrm{tr}\left[
\left(  \mathbf{\dot{h}}\cdot\mathbf{\boldsymbol{\sigma}}\right)  \left(
\mathbf{h}\cdot\mathbf{\boldsymbol{\sigma}}\right)  \right]  +\nonumber\\
&  +\frac{1}{2}\mathrm{tr}\left[  \left(  \mathbf{a}\cdot
\mathbf{\boldsymbol{\sigma}}\right)  \left(  \mathbf{h}\cdot
\mathbf{\boldsymbol{\sigma}}\right)  \left(  \mathbf{\dot{h}}\cdot
\mathbf{\boldsymbol{\sigma}}\right)  \right]  -\frac{1}{2}\mathrm{tr}\left[
\left(  \mathbf{a}\cdot\mathbf{\boldsymbol{\sigma}}\right)  \left(
\mathbf{\dot{h}}\cdot\mathbf{\boldsymbol{\sigma}}\right)  \left(
\mathbf{h}\cdot\mathbf{\boldsymbol{\sigma}}\right)  \right] \nonumber\\
&  =\frac{1}{2}\mathrm{tr}\left[  \left(  \mathbf{h}\cdot\mathbf{\dot{h}%
}\right)  \mathbf{1}+i\left(  \mathbf{h}\times\mathbf{\dot{h}}\right)
\cdot\mathbf{\boldsymbol{\sigma}}\right]  -\frac{1}{2}\mathrm{tr}\left[
\left(  \mathbf{\dot{h}}\cdot\mathbf{h}\right)  \mathbf{1}+i\left(
\mathbf{\dot{h}}\times\mathbf{h}\right)  \cdot\mathbf{\boldsymbol{\sigma}%
}\right]  +\nonumber\\
&  +\frac{1}{2}\mathrm{tr}\left\{  \left[  \left(  \mathbf{a}\cdot
\mathbf{h}\right)  \mathbf{1}+i\left(  \mathbf{a\times h}\right)
\cdot\mathbf{\boldsymbol{\sigma}}\right]  \left(  \mathbf{\dot{h}}%
\cdot\mathbf{\boldsymbol{\sigma}}\right)  \right\}  -\frac{1}{2}%
\mathrm{tr}\left\{  \left[  \left(  \mathbf{a}\cdot\mathbf{\dot{h}}\right)
\mathbf{1}+i\left(  \mathbf{a\times\dot{h}}\right)  \cdot
\mathbf{\boldsymbol{\sigma}}\right]  \left(  \mathbf{h}\cdot
\mathbf{\boldsymbol{\sigma}}\right)  \right\} \nonumber\\
&  =\mathbf{h}\cdot\mathbf{\dot{h}-\dot{h}}\cdot\mathbf{h+}\frac{1}%
{2}\mathrm{tr}\left\{  \left[  i\left(  \mathbf{a\times h}\right)
\cdot\mathbf{\boldsymbol{\sigma}}\right]  \left(  \mathbf{\dot{h}}%
\cdot\mathbf{\boldsymbol{\sigma}}\right)  \right\}  -\frac{1}{2}%
\mathrm{tr}\left\{  \left[  i\left(  \mathbf{a\times\dot{h}}\right)
\cdot\mathbf{\boldsymbol{\sigma}}\right]  \left(  \mathbf{h}\cdot
\mathbf{\boldsymbol{\sigma}}\right)  \right\} \nonumber\\
&  =\frac{i}{2}\mathrm{tr}\left\{  \left[  \left(  \mathbf{a\times h}\right)
\cdot\mathbf{\dot{h}}\right]  \mathbf{1}\right\}  -\frac{i}{2}\mathrm{tr}%
\left\{  \left[  \left(  \mathbf{a\times\dot{h}}\right)  \cdot\mathbf{h}%
\right]  \mathbf{1}\right\} \nonumber\\
&  =i\left[  \left(  \mathbf{a\times h}\right)  \cdot\mathbf{\dot{h}-}\left(
\mathbf{a\times\dot{h}}\right)  \cdot\mathbf{h}\right]  \text{,}%
\end{align}
that is,%
\begin{equation}
\left\langle \left[  \mathrm{H}\text{, }\mathrm{\dot{H}}\right]  \right\rangle
=i\left[  \left(  \mathbf{a\times h}\right)  \cdot\mathbf{\dot{h}-}\left(
\mathbf{a\times\dot{h}}\right)  \cdot\mathbf{h}\right]  \text{.}
\label{tyler2}%
\end{equation}
Then, using Eq. (\ref{tyler2}), we finally get%
\begin{equation}
\left\vert \left\langle \left[  \mathrm{H}\text{, }\mathrm{\dot{H}}\right]
\right\rangle \right\vert ^{2}=\left[  \left(  \mathbf{a\times h}\right)
\cdot\mathbf{\dot{h}-}\left(  \mathbf{a\times\dot{h}}\right)  \cdot
\mathbf{h}\right]  ^{2}\text{.} \label{tyler3}%
\end{equation}
As a side remark, we observe that writing $\mathbf{h}\left(  t\right)
=h(t)\hat{h}(t)$, we have $\mathbf{\dot{h}}\left(  t\right)  =\dot{h}%
(t)\hat{h}(t)+h(t)\partial_{t}\hat{h}(t)$. Therefore, we conclude that
$\left\vert \left\langle \left[  \mathrm{H}\text{, }\mathrm{\dot{H}}\right]
\right\rangle \right\vert ^{2}$ in Eq. (\ref{tyler3}) vanishes when the
magnetic field $\mathbf{h}\left(  t\right)  $ does not change in direction,
that is $\partial_{t}\hat{h}(t)=\mathbf{0}$ (i.e., $\mathbf{h}\left(
t\right)  $ and $\mathbf{\dot{h}}\left(  t\right)  $ are collinear). Moreover,
to double check the correctness of Eq. (\ref{tyler3}), we have used brute
force to calculate the RHS and LHS of Eq. (\ref{tyler3}) in an independent
manner. Indeed, setting%
\begin{equation}
\rho\overset{\text{def}}{\mathbf{=}}\frac{1}{2}\left(
\begin{array}
[c]{cc}%
a_{3} & a_{1}-ia_{2}\\
a_{1}+ia_{2} & -a_{3}%
\end{array}
\right)  \text{, \textrm{H}}\overset{\text{def}}{\mathbf{=}}\left(
\begin{array}
[c]{cc}%
h_{0}+h_{3} & h_{1}-ih_{2}\\
h_{1}+ih_{2} & h_{0}-h_{3}%
\end{array}
\right)  \text{, and \textrm{\.{H}}}\overset{\text{def}}{\mathbf{=}}\left(
\begin{array}
[c]{cc}%
d_{3} & d_{1}-id_{2}\\
d_{1}+id_{2} & -d_{3}%
\end{array}
\right)  \text{,}%
\end{equation}
with $\mathbf{a}\overset{\text{def}}{\mathbf{=}}\left(  a_{1}\text{, }%
a_{2}\text{, }a_{3}\right)  $, $\mathbf{h}\overset{\text{def}}{\mathbf{=}%
}\left(  h_{1}\text{, }h_{2}\text{, }h_{3}\right)  $, and $\mathbf{\dot{h}%
}\overset{\text{def}}{\mathbf{=}}\left(  d_{1}\text{, }d_{2}\text{, }%
d_{3}\right)  $, we have%
\begin{equation}
\left\vert \left\langle \left[  \mathrm{H}\text{, }\mathrm{\dot{H}}\right]
\right\rangle \right\vert ^{2}=\left(  2a_{1}d_{3}h_{2}-2a_{1}d_{2}%
h_{3}+2a_{2}d_{1}h_{3}-2a_{2}d_{3}h_{1}-2a_{3}d_{1}h_{2}+2a_{3}d_{2}%
h_{1}\right)  ^{2}\text{.}%
\end{equation}
Similarly, employing simple vector algebra, we get%
\begin{equation}
\left[  \left(  \mathbf{a\times h}\right)  \cdot\mathbf{\dot{h}-}\left(
\mathbf{a\times\dot{h}}\right)  \cdot\mathbf{h}\right]  ^{2}=\left(
2a_{1}d_{3}h_{2}-2a_{1}d_{2}h_{3}+2a_{2}d_{1}h_{3}-2a_{2}d_{3}h_{1}%
-2a_{3}d_{1}h_{2}+2a_{3}d_{2}h_{1}\right)  ^{2}\text{.}%
\end{equation}
Thus, we conclude that Eq. (\ref{tyler3}) is correct. With this final remark,
we end our discussion here.

\bigskip

\bigskip

\bigskip

\bigskip

\bigskip

\bigskip

\bigskip

\bigskip

\begin{thebibliography}{99}                                                                                               %


\bibitem {frey16}M. R. Frey, \emph{Quantum speed limits-primer, perspectives,
and potential future directions}, Quantum Inf. Process. \textbf{15}, 3919 (2016).

\bibitem {deffner17}S. Deffner and S. Campbell, \emph{Quantum speed limits:
From Heisenberg's uncertainty principle to optimal quantum control}, J. Phys.
A: Math. Theor. \textbf{50}, 453001 (2017).

\bibitem {mandelstam45}L. Mandelstam and Ig. Tamm, \emph{The uncertainty
relation between energy and time in non-relativistic quantum mechanics}, J.
Phys. USSR \textbf{9}, 249-254 (1945).

\bibitem {margo98}N. Margolus and L. B. Levitin, \emph{The maximum speed of
dynamical evolution}, Physica \textbf{D120}, 188 (1998); L. B. Levitin and T.
Toffoli, \emph{Fundamental limit on the rate of quantum dynamics: The unified
bound is tight}, Phys. Rev. Lett. \textbf{103}, 160502 (2009).

\bibitem {ole22}N. H\"{o}rnedal, D. Allan, and O. S\"{o}nnerborn,
\emph{Extensions of the Mandelstam-Tamm quantum speed limit to systems in
mixed states}, New J. Phys. \textbf{24}, 055004 (2022).

\bibitem {cai81}E. R. Caianiello, \emph{Is there a maximal acceleration?}
Lettere al Nuovo Cimento\textbf{\ 32}, 65 (1981).

\bibitem {cai82}E. R. Caianiello, S. De Filippo, G. Marmo, and G. Vilasi,
\emph{Remarks on the maximal-acceleration hypothesis}, Lettere al Nuovo
Cimento \textbf{34}, 112 (1982).

\bibitem {cai84}E. R. Caianiello, \emph{Maximal acceleration as a consequence
of Heisenberg's uncertainty relations}, Lettere al Nuovo Cimento \textbf{41},
370 (1984).

\bibitem {pati92}A. K. Pati, \emph{A note on maximal acceleration}, Europhys.
Lett. \textbf{18}, 285 (1992).

\bibitem {pati92b}A. K. Pati, \emph{On the maximal acceleration and the
maximal energy loss}, Il Nuovo Cimento \textbf{107B}, 895 (1992).

\bibitem {schot78}S. H. Schot, \emph{Jerk: The time rate of change of
acceleration},\ Am. J. Phys. \textbf{46}, 1090 (1978).

\bibitem {eager16}D. Eager, A.-M. Pendrill, and N. Reistad, \emph{Beyond
velocity and acceleration: Jerk, snap and higher derivatives}, Eur. J. Phys.
\textbf{37}, 065008 (2016).

\bibitem {masuda11}S. Masuda and K. Nakamura, \emph{Acceleration of adiabatic
quantum dynamics in electromagnetic fields}, Phys. Rev. \textbf{A84}, 043434 (2011).

\bibitem {masuda22}S. Masuda, J. Koenig, and G. A. Steele, \emph{Acceleration
and deceleration of quantum dynamics based on inter-trajectory travel with
fast-forward scaling theory}, Scientific Reports \textbf{12}, 10744 (2022).

\bibitem {nakamura16}A. Khujakulov and K. Nakamura, \emph{Scheme for
accelerating quantum tunneling dynamics}, Phys. Rev. \textbf{A93}, 022101 (2016)

\bibitem {larrouy20}A. Larrouy et \textit{al}., \emph{Fast navigation in a
large Hilbert space using quantum optimal control}, Phys. Rev. \textbf{X10},
02158 (2020).

\bibitem {saridis88}K. J. Kyriakopoulos and G. N. Saridis, \emph{Minimum jerk
path generation}, Proceedings of IEEE International Conference on Robotics and
Automation, vol. \textbf{1}, 364 (1988).

\bibitem {wise05}R. Shadmehr and S. P. Wise, \emph{The Computational
Neurobiology of Reaching and Pointing}, MIT Press (2005).

\bibitem {liu19}L. R. Liu et \textit{al}., \emph{Molecular assembly of
ground-state cooled single atoms}, Phys. Rev. \textbf{X9}, 021039 (2019).

\bibitem {matt23}A. J. Matthies et \textit{al}., \emph{Long-distance
optical-conveyor-belt transport of ultracold} $^{133}$\textrm{Cs }%
\emph{and}\textrm{\ }$^{87}$\textrm{Rb }\emph{atoms}, Phys. Rev.
\textbf{A109}, 023321 (2024).

\bibitem {pati23}A. K. Pati, \emph{Quantum acceleration limit},
arXiv:quant-ph/2312.00864 (2023).

\bibitem {paul24A}P. M. Alsing and C. Cafaro, \emph{From the classical
Frenet-Serret apparatus to the curvature and torsion of quantum-mechanical
evolutions. Part I. Stationary Hamiltonians}, Int. J. Geom. Methods Mod. Phys.
\textbf{21}, 2450152 (2024).

\bibitem {paul24B}P. M. Alsing and C. Cafaro, \emph{From the classical
Frenet-Serret apparatus to the curvature and torsion of quantum-mechanical
evolutions. Part II. Nonstationary Hamiltonians}, Int. J. Geom. Methods Mod.
Phys. \textbf{21}, 2450151 (2024).

\bibitem {alsing24}P. M. Alsing and C. Cafaro, \emph{Upper limit on the
acceleration of a quantum evolution in projective Hilbert space}, Int. J.
Geom. Methods Mod. Phys. \textbf{21}, 2440009 (2024).

\bibitem {heisenberg}W. Heisenberg, \emph{\"{U}ber den anschaulichen Inhalt
der quanten theoretischen kinematik und mechanik}, Z. Physik \textbf{43},172 (1927).

\bibitem {kennard}E. H. Kennard, \emph{Zur Quanten mechanik einfacher
Bewegungstypen}, Z. Physik \textbf{44},326 (1927).

\bibitem {robertson}H. P. Roberston,\emph{The uncertainty principle}, Phys.
Rev. \textbf{34},163 (1929).

\bibitem {schrodinger}E. Schr\"{o}dinger,\emph{\ Zum Heisenbergschen
\emph{Unsch\"{a}}rfeprinzip}, Sitzungsberichte der Preussischen Akademie der
Wissenschaften, Physikalisch-mathematische Klasse \textbf{14}, 296 (1930).

\bibitem {englert}B.-G. Englert, \emph{Uncertainty relations revisited}, Phys.
Lett. \textbf{A494}, 129278 (2024).

\bibitem {nakahara}M. Nakahara, \emph{Geometry, Topology, and Physics},
Institute of Physics Publishing Ltd (2003).

\bibitem {eguchi80}T. Eguchi, P. B. Gilkey, and A. J. Hanson,
\emph{Gravitation, gauge theories, and differential geometry}, Phys. Rep.
\textbf{66}, 213 (1980).

\bibitem {bohm91}A. Bohm, L. J. Boya, and B. Kendrick, \emph{Derivation of the
geometric phase}, Phys. Rev. \textbf{A43}, 1206 (1991).

\bibitem {crell09}A. Uhlmann and B. Crell, \emph{Geometry of state spaces},
Lecture Notes in Physics \textbf{768}, 1 (2009).

\bibitem {provost80}J. P. Provost and G. Vallee, \emph{Riemannian structure on
manifolds of quantum states}, Commun. Math. Phys. \textbf{76}, 289 (1980).

\bibitem {mukunda93}N. Mukunda and R. Simon, \emph{Quantum kinematic approach
to the geometric phase I. General Formalism}, Annals of Physics\textbf{\ 228},
205 (1993).

\bibitem {anandan90}J. Anandan and Y. Aharonov, \emph{Geometry of quantum
evolution}, Phys. Rev. Lett. \textbf{65}, 1697 (1990).

\bibitem {sam94}S. L. Braunstein and C. M. Caves, \emph{Statistical distance
and the geometry of quantum states}, Phys. Rev. Lett. \textbf{72}, 3439 (1994).

\bibitem {uzdin12}R. Uzdin, U. G\"{u}nther, S. Rahav, and N. Moiseyev,
\emph{Time-dependent Hamiltonians with 100\% evolution speed efficiency}, J.
Phys. A: Math. Theor. \textbf{45}, 415304 (2012).

\bibitem {sakurai85}J. J. Sakurai, \emph{Modern Quantum Mechanics}, Addison
Wesley Publishing Company, Inc. (1985).

\bibitem {hall13}B. C. Hall, \emph{Quantum Theory for Mathematicians},
Springer Science$+$Business Media New York (2013).

\bibitem {cafaro24A}C. Cafaro, L. Rossetti, and P. M. Alsing, \emph{Curvature
of quantum evolutions for qubits in time-dependent magnetic fields},
arXiv:quant-ph/2408.14233 (2024).

\bibitem {leo24}L. Rossetti, C. Cafaro, and P. M. Alsing,\emph{ Quantifying
deviations from shortest geodesic paths together with waste of energy
resources for quantum evolutions on the Bloch sphere,}\textbf{ }%
arXiv:quant-ph/2408.14230 (2024).

\bibitem {jakob01}L. Jakobczyk and M. Siennicki, \emph{Geometry of Bloch
vectors in two-qubit system}, Phys. Lett. \textbf{A286}, 383 (2001).

\bibitem {kimura03}G. Kimura, \emph{The Bloch vector for }$N$\emph{-level
systems}, Phys. Lett. \textbf{A314}, 339 (2003).

\bibitem {krammer08}R. A. Bertlmann and P. Krammer, \emph{Bloch vectors for
qudits}, J. Phys. A: Math. Theor. \textbf{41}, 235303 (2008).

\bibitem {kurzy11}P. Kurzynski, \emph{Multi-Bloch vector representation of the
qutrit}, Quantum Inf. Comp. \textbf{11}, 361 (2011).

\bibitem {xie20}J. Xie et \textit{al}., \emph{Observing geometry of quantum
states in a three-level system}, Phys. Rev. Lett. \textbf{125}, 150401 (2020).

\bibitem {siewert21}C. Eltschka, M. Huber, S. Morelli, and J. Siewert,
\emph{The shape of higher-dimensional state space: Bloch-ball analog for a
qutrit}, Quantum \textbf{5}, 485 (2021).

\bibitem {gamel16}O. Gamel,\emph{\ Entangled Bloch spheres: Bloch matrix and
two-qubit state space}, Phys. Rev. \textbf{A93}, 062320 (2016).

\bibitem {bures69}D. Bures, \emph{An extension of Kakutani's theorem on
infinite product measures to the tensor product of semifinite }$\omega^{\ast}
$\emph{-algebras}, Trans. Amer. Math.\ Soc. \textbf{135}, 199 (1969).

\bibitem {uhlman76}A. Uhlmann, \emph{The \textquotedblleft transition
probability\textquotedblright\ in the state space of a }$\ast$\emph{-algebra},
Rep. Math. Phys. \textbf{9}, 273 (1976).

\bibitem {hubner92}M. H\"{u}bner, \emph{Explicit computation of the Bures
distance for density matrices}, Phys. Lett. \textbf{A163}, 239 (1992).

\bibitem {karol06}I. Bengtsson and K. Zyczkowski, \emph{Geometry of Quantum
States}, Cambridge University Press (2006).

\bibitem {erik20}E. Sj\"{o}qvist, \emph{Geometry along evolution of mixed
quantum states}, Phys. Rev. Research \textbf{2}, 013344 (2020).

\bibitem {alsingpra23}P. M. Alsing, C. Cafaro, O. Luongo, C. Lupo, S. Mancini,
and H. Quevedo, \emph{Comparing metrics for mixed quantum states: Sj\"{o}qvist
and Bures}, Phys. Rev. \textbf{A107}, 052411 (2023).

\bibitem {chi24}X.-Y. Hou et \textit{al}., \emph{Local geometry and quantum
geometric tensor of mixed states}, Phys. Rev. \textbf{B110}, 035144 (2024).

\bibitem {silva21}H. Silva, B. Mera, and N. Paunkovic, \emph{Interferometric
geometry from symmetry-broken Uhlmann gauge group with applications to
topological phase transitions}, Phys. Rev. \textbf{B103}, 085127 (2021).

\bibitem {silva21B}H. V. da Silva, \emph{Quantum information geometry and
applications}, MS\ Thesis in Engineering Physics, IT Lisboa (2021).

\bibitem {mera22}B. Mera, N. Paunkovic, S. T. Amin, and V. R. Vieira,
\emph{Information geometry of quantum critical submanifolds: Relevant,
marginal, and irrelevant operators}, Phys. Rev. \textbf{B106}, 155101 (2022).

\bibitem {campaioli19}F. Campaioli, W. Sloan, K. Modi, and F. A. Pollock,
\emph{Algorithm for solving unconstrained unitary quantum brachistochrone
problems}, Phys. Rev. \textbf{A100}, 062328 (2019).

\bibitem {cafaroQR}C. Cafaro and P. M.\ Alsing, \emph{Minimum time for the
evolution to a nonorthogonal quantum state and upper bound of the geometric
efficiency of quantum evolutions}, Quantum Reports \textbf{3}, 444 (2021).

\bibitem {carloprd22}C. Cafaro and P. M. Alsing, \emph{Complexity of pure and
mixed qubit geodesic paths on curved manifolds}, Phys. Rev.\textbf{\ D106},
096004 (2022).

\bibitem {carlopre22}C. Cafaro, S. Ray, and P. M. Alsing, \emph{Complexity and
efficiency of minimum entropy production probability paths from quantum
dynamical evolutions}, Phys. Rev. \textbf{E105}, 034143 (2022).

\bibitem {carlocqg23}C. Cafaro and P. M. Alsing, \emph{Qubit geodesics on the
Bloch sphere from optimal-speed Hamiltonian evolutions}, Class. Quantum Grav.
\textbf{40}, 115005 (2023).

\bibitem {saito23}T. Van Vu and K.\ Saito, \emph{Thermodynamic unification of
optimal transport: Thermodynamic uncertainty relation, minimum dissipation,
and thermodynamic speed limits}, Phys. Rev. \textbf{X13}, 011013 (2023).

\bibitem {araki23}T. Araki, F. Nori, and C. Gneiting, \emph{Robust quantum
control with disorder-dressed evolution}, Phys. Rev. \textbf{A107}, 032609 (2023).

\bibitem {hamazaki23}R. Hamazaki,\emph{\ Limits to fluctuation dynamics},
Communications Physics\textbf{\ 7}, 361 (2024).

\bibitem {hamazaki22}R. Hamazaki, \emph{Speed limits for macroscopic
transitions}, PRX Quantum \textbf{3}, 020319 (2022).

\bibitem {flores23}C. Chryssomalakos et \textit{al}., \emph{Curves in quantum
state space, geometric phases, and the brachistophase}, J. Phys. A: Math.
Theor. \textbf{56}, 285301 (2023).

\bibitem {flores24}C. Chryssomalakos et \textit{al}., \emph{Speed excess and
total acceleration: A kinetical approach to entanglement},
arXiv:quant-ph/2401.17427 (2024).

\bibitem {gibilisco08}P. Gibilisco, D. Imparato, and T. Isola, \emph{A
Robertson-type uncertainty principle and quantum Fisher information}, Linear
Algebra Appl. \textbf{428}, 1706 (2008).

\bibitem {maccone14}L. Maccone and A. K. Pati, \emph{Stronger uncertainty
relations for all incompatible observables}, Phys. Rev. Lett. \textbf{113},
260401 (2014).

\bibitem {nielsen00}M. A. Nielsen and I. L. Chuang, \emph{Quantum Computation
and Quantum Information}, Cambridge University Press (2000).

\bibitem {polak11}E. G. Rieffel and W. H. Polak, \emph{Quantum Computing: A
Gentle Introduction}, MIT Press (2011).

\bibitem {hidary19}D. Hidary, \emph{Quantum Computing: An Applied Approach},
Springer (2019).
\end{thebibliography}
\end{document}